\documentclass[prb,preprint,amsmath,amssymb]{revtex4}
\usepackage{color}
\begin{document}
\author{Kevin Leung$^*$}
\affiliation{Sandia National Laboratories, MS 1415, Albuquerque, NM 87185\\
$^*${\tt kleung@sandia.gov}}
\date{\today}
\title{First Principles, Explicit Interface Studies of Oxygen Vacancy
and Chloride in Alumina Films for Corrosion Applications}

\input epsf.sty
 
\begin{abstract}

Pitting corrosion is a much-studied and technologically relevant subject.
However, the fundamental mechanisms responsible for the breakdown of the 
passivating oxide layer are still subjects of debate.  Chloride
anions are known to accelerate corrosion; relevant hypotheses include
Cl insertion into positively charged oxygen vacancies in the oxide film,
and Cl adsorption on passivating oxide surfaces, substituting for surface
hydroxyl groups.  In this work, we conduct large-scale first principles
modeling of explicit metal/Al$_2$O$_3$ interfaces to investigate the
energetics and electronic structures associated with these hypotheses.
The explicit interface models allow electron transfer
that mimics electrochemical events, and the establishment of the relation
between atomic structures at different interfaces and the electronic band
alignment.  For multiple model interfaces, we find that doubly charged oxygen
vacancies, which are key ingredients of the point defect model (PDM) often
used to analyze corrosion data, can only occur in the presence of a
potential gradient that raises the voltage.  Cl$^-$ insertion into oxide
films can be energetically favorable in some oxygen vacancy sites, depending
on the voltage.  We also discuss the challenges associated with explicit DFT
modeling of these complex interfaces.

\end{abstract}
 
\maketitle
 
\section{Introduction}
\label{intro}

Corrosion is known to cost billions to industry per year.\cite{corrbook}
Extensive field work and benchtop experiments have been devoted to
understanding and mitigating corrosion effects.  Despite this, aspects of 
the fundamental mechanisms responsible for pitting corrosion, which involves
localized breakdown of passivating oxides on metal surfaces, remain debated.

\begin{figure}
\centerline{\hbox{  \epsfxsize=3.00in \epsfbox{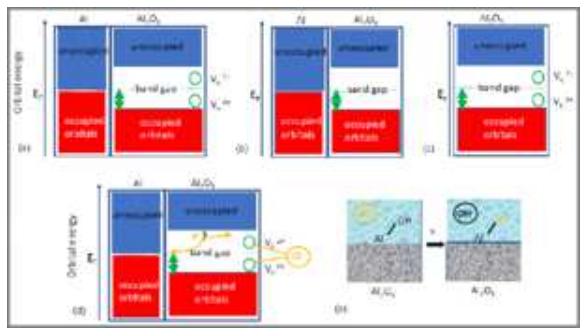} }}
\caption[]
{\label{fig1} \noindent
Schematics of key science questions addressed.  (a) The net charge in
V$_{\rm O}$ depends on the energy level of the defect orbital relative to
$E_{\rm F}$.  This can be computed in two steps: (b) no vacancy and then (c) no
interface.  If there is a electric field in the oxide film and/or a net charge
transfer in the electrode, however, (a) must be used.  (d) Cl$^-$ insertion
into oxygen vacancies.  (e) Cl$^-$ substitution for a OH$^-$ surface group.
}
\end{figure}

In this work, we apply Density Functional Theory (DFT) calculations to
investigate the defect electronic structure and the energetics of oxygen
vacancies and Cl$^-$ anions in model oxide films in direct contact with aluminum
metal.  These oxide films are known to be a few nanometers
thick.\cite{thickness}
Our models are geared towards atmospheric corrosion conditions but
our predictions are also relevant to aluminum immersed in aqueous electrolytes.
We adopt computationally costly DFT models with explicit metal/oxide
interfaces to allow electron transfer between defects in the passivating oxide
film and the metal.  This allows us to mimic electrochemical (electron
transfer) events and generate finite electric fields in the oxide.  Both will
be shown to be crucial for making connections to continuum models which have
been used to analyze corrosion.\cite{macdonald81a,macdonald81b}
We focus on two hypotheses postulated in the literature related to pitting
initiation via breakdown in alumina surface films.  One revolves around the
key role of positively charged oxygen vacancies (V$_{\rm O}$).  The other,
which we touch on more briefly, 
concerns Cl$^-$ substitution of surface OH$^-$ groups.  See Fig.~\ref{fig1}.

Positively charged V$_{\rm O}$ are key ingredients of the point defect model
(PDM)\cite{macdonald81a,macdonald81b,macdonald15,macdonald16a,macdonald16b,dft_ingress,macdonald04}
widely used to analyze time-dependent corrosion behavior.
Cl$^-$ is an impurity well-known to accelerate pitting corrosion in many
metals.\cite{cl,white,natishan2014,natishan2017,natishan2000,mccafferty_review,marcus_review}
Cl$^-$ is more detrimental than other halide anions.\cite{macdonald16a}
Cl$^-$ insertion into V$_{\rm O}^{2+}$ has been suggested to play a main
role in corrosion via a Cl$^-$ insertion \color{black}
reaction:\cite{macdonald81b} \color{black}
\begin{equation}
{\rm V}_{\rm O}^{n+} + {\rm Cl}^- (aq) \rightarrow
{\rm Cl}_{\rm Vo}^{m+} + (m-n-1) e^- ,  \label{eqcl1}
\end{equation}
where ``(aq)'' denotes aqueous phase.  The electron ($e^-$) released
indicates that Eq.~\ref{eqcl1} is electrochemical in nature, and is therefore
voltage-dependent.  Indeed, in \color{black} benchtop \color{black}
experiments, applying voltages above
a threshold is linked to the onset of pitting corrosion.\cite{laycock,science}
In aluminum metal immersed in aqueous electrolytes, the threshold appears to
be at $\Phi_{\rm ext}$$>$-0.5~V vs.~standard hydrogen electrode (SHE),
depending on the salt concentration and pH.\cite{pittingvoltage,read} 
Cl$^-$ implantation and other experiments have been performed to interrogate
this hypothesis.\cite{sullivan_cl,ion_implant,wood} 

To model $e^-$ transfer effects on corrosion requires a quantum mechanical
treatment of valence $e^-$, with DFT being a standard compromise between
accuracy and computational cost.  Our first step is to apply DFT models with
explicit Al$|$Al$_2$O$_3$ interfaces to investigate the net charge in oxygen
vacancies, and examine under what conditions V$_{\rm O}^{2+}$ ($n$=2) exists.
Fig.~\ref{fig1} illustrates this key issue: the net charges of defects like
O-vacancies cannot be manually assigned, but are determined by the the vacancy
orbital energy level relative to the Fermi level ($E_{\rm F}$).  

Previous DFT work that has studied the Al metal/Al$_2$O$_3$
interface\cite{bredas,kleinman,nieminen,valone,smith2000,siegel2002,wang,liu2014,eremeev,marks,marks1,costa_o2,mira,al2o3_defect}  
has not simultaneously considered defects like oxygen vacancies.  
While oxygen vacancies in Al$_2$O$_3$ have been examined using
DFT,\cite{al2o3_ovac0,al2o3_ovac1,al2o3_ovac2,al2o3_ovac3,al2o3_ovac4,hine}
most such calculations are conducted in the absence of metallic Al, which
defines the Fermi level ($E_{\rm F}$), in the simulation cell.
$E_{\rm F}$ is not well defined in insulators like Al$_2$O$_3$ without a
metal electrode because $E_{\rm F}$ 
should be pinned at $p$- or $n$-type defects seldom explicitly depicted in DFT
simulation cells.  As such, $E_{\rm F}$ has generally been treated as a free
parameter.  Furthermore, oxides modeled using periodic boundary conditions
in all three spatial dimensions cannot support an electric field -- unless the
Berry's phase approach is used.\cite{berry} Therefore oxide calculations
in the absence of an metal electrode and an explicit metal/oxide interface
assume a zero-field, ``flat-band'' approximation.\cite{dabo1}  In addition,
at metal/oxide interfaces, a contact potential at the metal/oxide interface
exists; its value can be up to 2-3~V,\cite{marks,pccp,costa20} and its effect
can only be captured if an explicit metal/oxide interface exists in the
simulation cell.
Finally, there are at least two interfaces, namely metal/oxide and
oxide/electrolyte (Fig.~\ref{fig1}).  These interfaces both contribute to the
overall voltage measured in experiments, but how these interfacial structures 
separately affect the electronic band alignment is not well established.

Here we adopt simulation cells with both an explicit Al$|$Al$_2$O$_3$ interface
and an explicit V$_{\rm O}$.  Using several interface models, we show that
uncharged V$_{\rm O}$ \color{black} in the oxide film \color{black}
is most likely the norm under flat-band
conditions; V$_{\rm O}^{2+}$ are found to occur only in the presence of
electric fields in the oxide film.  In other words, aluminum metal shows less
tendency to support V$_{\rm O}^{2+}$ \color{black} in the oxide film 
\color{black} than expected, likely because it is among the more
electronegative metals.  Indeed, previous DFT modeling of amorphous
Al$_2$O$_3$ has suggested overall negative charging due to multiple
defects.\cite{shluger2,shluger1}  We also investigate the energetics of
Cl$^-$ insertion into V$_{\rm O}$ and V$_{\rm O}^{2+}$ (Fig.~\ref{fig1}d), and
show that insertion is favorable under certain conditions.  
Our work arguably represents an important step towards parameterization of the
PDM model using first principles predictions.  \color{black} Note that one of
the few DFT studies with both explicit metal/oxide interfaces and
vacancies concerns Cr$_2$O$_3$ and CrOOH at Cr interfaces.
The authors there assume that the O-vacancies created are doubly positiviely
charged, but do not confirm it.  We propose that their V$_{\rm O}$ charge
assignment may need to be re-examined.\cite{cr_interface}  \color{black}

For completeness, we also examine another main hypothesis of aluminum
oxide film depassivation which concerns Cl$^-$ substituting for
OH$^-$ on Al$_2$O$_3$ surfaces (Fig.~\ref{fig1}e).  Specifically,
\begin{equation}
{\rm Al}_2{\rm O}_{3-n/2}{\rm (OH)}_n (s) + {\rm Cl}^- (aq) \rightarrow
{\rm Al}_2{\rm O}_{3-n/2}{\rm (OH)}_{n-1}{\rm Cl} (s) + {\rm OH}^- (aq) .
		\label{eqcl2}
\end{equation}
X-ray Photoemission Spectroscopy (XPS) measurements supported by DFT
calculations of core electron spectra have made a convincing case that
this substitution, without electron transfer,
takes place on oxide-passivated Al metal
surfaces.\cite{natishan1999,natishan2002,natishan2011,marcus2020} 
Computationally, fluorination of AlOH groups using HF gas has been
predicted to be energetically favorable.\cite{fluorination}  Cl$^-$ 
substitution for surface OH$^-$ and insertion into oxide films have also
been examined using DFT methods in Al$_2$O$_3,\cite{cl-al2o3}
$NiO,\cite{cl-nio} $\alpha$-Cr$_2$O$_3$,\cite{cl-cr2o3a,cl-cr2o3b} 
$\alpha$-Fe$_2$O$_3$,\cite{cl-fe2o3a,cl-fe2o3b} and other oxides.
However, Eq.~\ref{eqcl2} appears at odds with geochemistry research showing
that Cl$^-$ is less effective than other anions like SO$_4^{2-}$ to cause
dissolution of aluminum oxy-hydroxide.\cite{kolics1,kolics2,dietzel} In this
work, we report DFT energetics associated with Cl$^-$$\rightarrow$OH$^-$ on
the $\alpha$-Al$_2$O$_3$ (0001) surface using parameters consistent with
those we apply to examine Cl$^-$ insertion into V$_{\rm O}$ inside the oxide
film, so that a comparison of mechanisms can be made.

The passivating alumina films on Al metal surfaces are known to be
amorphous.\cite{persson} We apply $\alpha$-Al$_2$O$_3$ as a model for
the oxide film because it has interfaces which are better characterized;
much of the DFT work in the has literature also adopted $\alpha$-Al$_2$O$_3$.
Grain boundaries in $\alpha$-Al$_2$O$_3$, which may
be better local approximations of amorphous oxides, are also considered.
Our models and DFT calculations focus on atmospheric corrosion relevant to
electronics bond wire degradation,\cite{sand3008} where Al metal, Au, and
Al-Al intermetallics are all present.  Despite this, we draw on perspectives,
concepts, and pitting onset voltage dependences developed for Al corrosion
in aqueous environments.

This work focuses on the solid interface and defect aspects.  We will show
that quantitative comparison of predicted voltages with experimental
measurements may require future accurate treatment of the aqueous electrolyte.
Other calculations in the literature relevant to corrosion deal with DFT
screening of organic coating molecules\cite{costa_inhibitor,taylor_inhibitor}
and MD simulations of water- and Cl$^-$ adsorption surface sites in aqueous
environments.\cite{criscenti,costa,valero,klein18,klein19}  In the future,
these elements can be added to explicit metal/oxide interface modeling, either
in the same simulation cells or indirectly, via parameter passing protocols.

\begin{figure}
\centerline{\hbox{ (a) \epsfxsize=2.50in \epsfbox{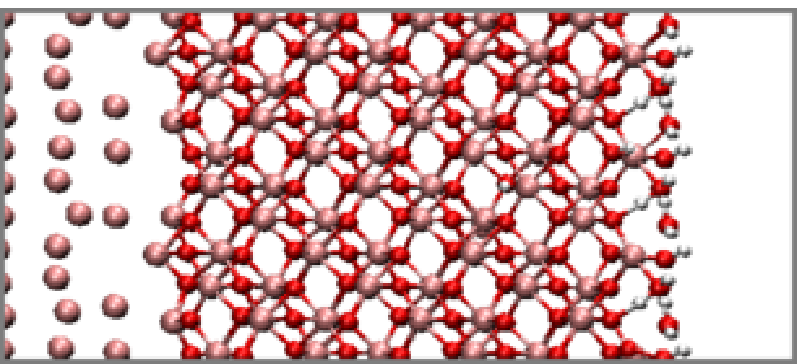} }}
\centerline{\hbox{ (b) \epsfxsize=2.50in \epsfbox{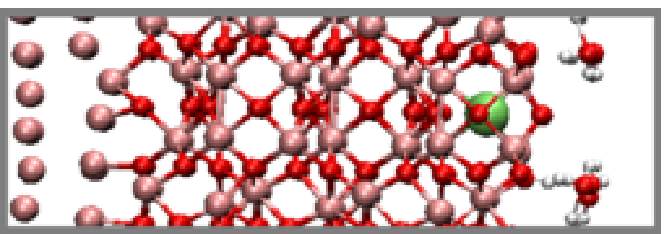} }}
\centerline{\hbox{ (c) \epsfxsize=2.50in \epsfbox{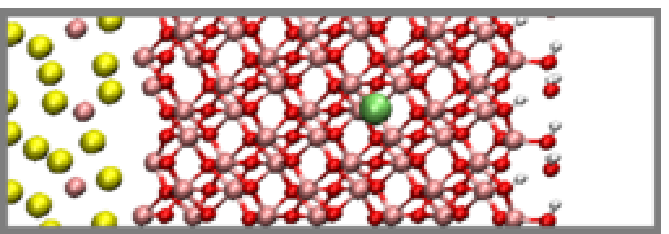} }}
\caption[]
{\label{fig2} \noindent
Interfacial systems considered in this work.
(a) $\alpha$-Al$_2$O$_3$ (0001) on Al(111);
(b) $\alpha$-Al$_2$O$_3$ with grain boundary on Al(111);
(c) $\alpha$-Al$_2$O$_3$ (0001) on Au(111).
Pink, yellow, red, white, and green represent Al, Au, O, and H atoms,
respectively.  Cl$^-$ insertion are shown in two cases.  The $+z$ direction
points from the metal towards the oxide.
}
\end{figure}

\section{Method and Models}

\subsection{DFT Details}
\label{dft}

Most DFT calculations in this work are conducted under T=0~K ultra-high vacuum
(UHV) condition, using periodically replicated simulation cells and the Vienna
Atomic Simulation Package (VASP) version 5.4.\cite{vasp1,vasp1a,vasp2,vasp3}
A 400~eV planewave energy cutoff and a 10$^{-4}$~eV convergence criterion
are enforced.  Most calculations apply the PBE functional.\cite{pbe}  In
some cases, HSE06 is used for spot check.\cite{hse06a,hse06b,hse06c}   
Spin-polarization is turned off because a net spin is found to only
accumulate in the metallic region, and it barely affects the total energy.

\begin{table}\centering
\begin{tabular}{l|l|l|r|r|l|r} \hline
metal & oxide & gb & dimensions & stoichiometry & $k$-sampling & 
		Figure \\ \hline
Al(111) & $\alpha$ (001) & no &  4.81$\times$8.33$\times$52.00 & 
  (Al$_{34})$Al$_{36}$O$_{54}($H$_2$O)$_{6}$ & 3$\times$2$\times$1  
				& Fig.~\ref{fig2}a \\
Al(111) & $\alpha$ (001) & no &  14.43$\times$16.663$\times$50.00 & 
(Al$_{204})$Al$_{216}$O$_{324}($H$_2$O$)_{36}$ & 2$\times$2$\times$1  & 
				Fig.~\ref{fig2}a \\
Al(111) & $\alpha$ & yes & 14.42$\times$26.87$\times$40.00 & 
	(Al$_{216})$Al$_{264}$O$_{420}$(H$_2$O)$_{24}$ & 2$\times$1$\times$1  
		& Fig.~\ref{fig2}b \\
Au(111) & $\alpha$ (001) & no & 14.43$\times$16.66$\times$50.00 & 
	(Au$_{180}$Al$_{24})$Al$_{216}$O$_{324}$(H$_2$O)$_{24}$ 
			& 2$\times$2$\times$1 & Fig.~\ref{fig2}c \\
none & $\alpha$ (bulk) & no & 14.43$\times$12.45$\times$13.08 & 
	Al$_{108}$O$_{162}$ & 2$\times$2$\times$2  & NA \\
none & $\alpha$ (bulk) & no & 9.58$\times$8.30$\times$13.08 & 
	Al$_{48}$O$_{72}$ & 2$\times$2$\times$2  & NA \\
\hline
\end{tabular}
\caption[]
{\label{table1} \noindent
\color{black}
Computational details of representative, baseline simulation cells.  The larger
of the two ``Fig.~\ref{fig2}a'' cell is used when inserting a V$_{\rm O}$.
Simulation cells which are 2$\times$1 and 2$\times$2 expansions in the lateral
dimensions are also considered.  Dimensions are in units of \AA$^3$.  
Brackets separate the metal and water zones from the oxide region.
All oxide films are stoichiometric Al$_2$O$_3$, and initially, before
selective deprotonation, integer numbers of H$_2$O molecules are added.
There are excess number of metal atoms at the interface, so the metal
zone contains non-integer number of metal layers.
\color{black}
}
\end{table}

Representative, baseline simulation cells are listed in Table~\ref{table1}.
The models with explicit interfaces are depicted in Fig.~\ref{fig2};  they
contain a vacuum region on top (in the $+z$ direction), and a Al(111) metal
slab at the bottom, with the bottom-most layer of metal atoms kept frozen.  We
also consider an Au(111) example (Fig.~\ref{fig2}c), because Au is present in
microelectronics bond pads,\cite{sand3008} and because it represents a metal
with a higher work function and an interesting comparison with Al.  The metal
films are covered with $\alpha$-Al$_2$O$_3$.  The (0001) facet
is in contact with the metal and is exposed to vacuum, except for
the case with $\Sigma_3$ (001) grain boundaries which run parallel to the
(0001) plane (Fig.~\ref{fig2}b).\cite{gb}  The $x$-dimension of this grain
boundary simulaton cell is kept at the bulk crystal values while the
$y$-dimension, perpendicular to the grain boundaries, is obtained by optimizing
a simulation cell with only the oxide present.  This mimics 
isolated grain boundaries, which should not affect the lattice dimension
parallel to it in a crystal of infinite size.

Next we discuss the details of the metal/oxide interfacial structures, which
have not been elucidated in experiments for all metal and/or oxide facets.
Even the widely adopted ``Al-termination'' configuration for the most-studied
interface ($\alpha$-Al$_2$O$_3$(0001)/Al(111)) has been constructed with a fixed
stoichiometry and does not allow the number of Al atoms in the surface unit
cell to vary, e.g., at constant Al chemical potential via a Grand Canonical
Ensemble simulation, which would be the most rigorous approach.  Our
Al(111)/$\alpha$-Al$_2$O$_3$ structure (Fig.~\ref{fig2}a) is based
on the ``FCC'' interface structure\cite{siegel2002} but adopts a doubled
surface unit cell of lateral dimension 4.81$\times$8.33~\AA$^2$.\cite{bredas}
For this cell, we conduct a 1.7~ps {\it ab initio} molecular dynamics (AIMD) at
T=400~K followed by quenching to T=0~K to equilibrate the 
Al(111)$|$$\alpha$-Al$_2$O$_3$(001) interface before optimizing the atomic
configuration.\cite{pccp}  In the appendix, we use a combinatorial approach to
compare the structure with other interfacial models investigated in the
literature.\cite{bredas,kleinman,nieminen,valone,smith2000,siegel2002,wang,liu2014} 
As Ref.~\onlinecite{bredas} points out, variations in the interfacial structure
can change the work function by $<$0.5~eV.

A similar AIMD procedure is applied to construct the
Au(111)$|$$\alpha$-Al$_2$O$_3$(001) interface (Fig.~\ref{fig2}c); this model
is meant to mimic Al$_2$O$_3$ grown on AlAu$_4$ intermetallic
surfaces.\cite{alau4a,alau4b} \color{black} The interface model with grain
boundaries (Fig.~\ref{fig2}b) is \color{black}
too large and costly to permit AIMD pre-equilibration.
For this model we displace the oxide and metal in the lateral directions and
show that the displacements only weakly affect the work function in the
appendix.  $\gamma$-Al$_2$O$_3$ surfaces have been the subjects of DFT studies,
but more than one atomic interpretations of the structure have been
reported.\cite{gal2o3_1,costa_o2,costa,valero}  $\gamma$-Al$_2$O$_3$ may be
\color{black} a \color{black}
better model for amorphous oxide, and will be considered in future work.

The number of water molecules residing on the outer surface of the oxide film
depends on the humidity.  Most existing computational work on explicit
Al$|$Al$_2$O$_3$ interfaces\cite{wang,bredas,persson} have omitted H$_2$O
molecules.  We include a submonolayer of H$_2$O not only because of the finite
humidity, but also because on some Al$_2$O$_3$ surface facets, the oxides
can exhibit metallic behavior at their outer (vacuum) surfaces in the absence
of H$_2$O or other solvent molecules when the inner oxide surfaces are in
contact with \color{black} electronegative metal.\cite{pccp} \color{black} As
an approximation of the finite temperature water film, we consider T=0~K
conditions, add sufficient H$_2$O molecules so each surface Al cation is
coordinated to the O-atom of 1-3 water molecules, and optimize the atomic
configurations.  Some H$_2$O molecules spontaneously react with the surface
and dissociate into two AlOH groups.  
\color{black}
In Fig.~\ref{fig2}a, each exposed Al$^{3+}$ on the surface is coordinated to
3~H$_2$O molecules, one of which is spontaneously hydrolysed, i.e., it
transfers a proton to a surface O$^{2-}$ otherwise bonded to 3 Al$^{3+}$.  
In Fig.~\ref{fig2}c, each exposed 3-coordinated Al$^{3+}$ is coordinated
to \color{black} an \color{black}
H$_2$O molecule, which is hydrogen-bonded to another adsorbed,
hydrolysed H$_2$O (i.e., OH group).  50\% of the H$_2$O are hydrolysed.
In Fig.~\ref{fig2}b, each exposed Al$^{3+}$ on the surface is coordinated to
one~OH from \color{black} an \color{black}
H$_2$O which has donated \color{black} an \color{black}
H$^+$ to a surface O$^{2-}$; this
newly created OH group is coordinated to another H$_2$O molecule via 
hydrogen bonding.  37.5~\% of the adsorbed H$_2$O molecules are hydrolysed.
\color{black} Like at the metal/oxide interface, we do
not claim to have found the most stable H$_2$O adsorption configuration on
all Al$_2$O$_3$ surface facets; some of our H$_2$O configurations may be
metastable.  Instead, we vary the interfacial structures and examine the
differences in the appendix, where we show that a change in the H$_2$O
structure can give rise to a $>$1~eV change in the work function
and a similar global shift in the band structure,\cite{campbell1} but this
does not strongly affect the metal/oxide band alignment.
The different water terminations
in Fig.~\ref{fig2}a and Fig.~\ref{fig2}c help illustrate this point.

\begin{figure}
\centerline{\hbox{ (a) \epsfxsize=1.50in \epsfbox{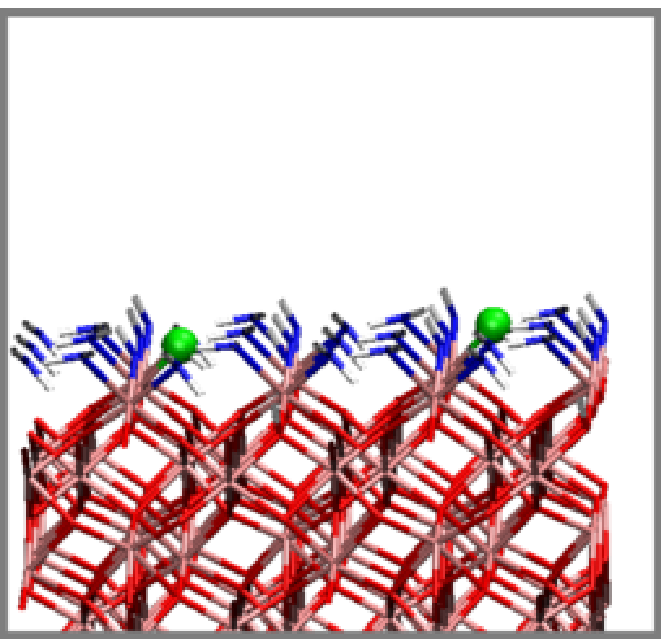}
		   (b) \epsfxsize=1.50in \epsfbox{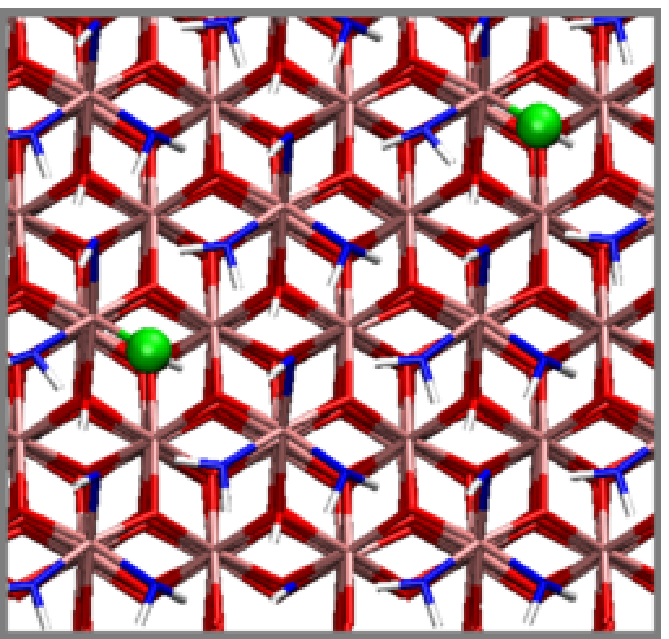} }}
\centerline{\hbox{ (c) \epsfxsize=1.50in \epsfbox{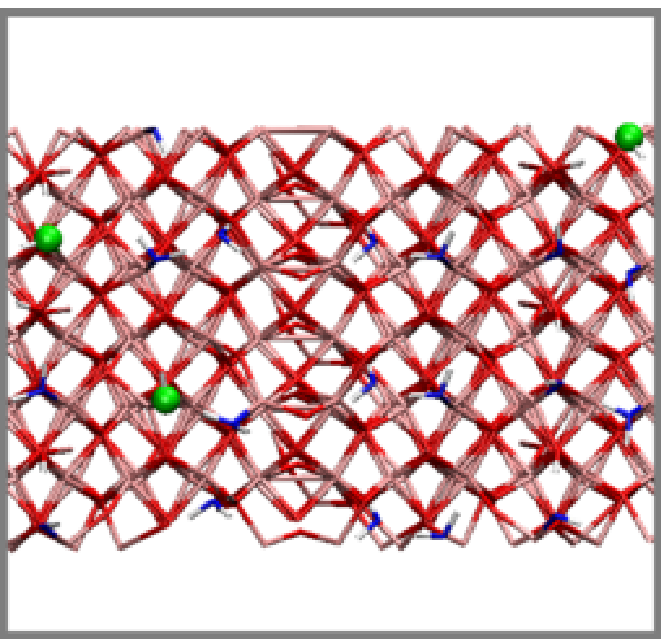}
		   (d) \epsfxsize=1.50in \epsfbox{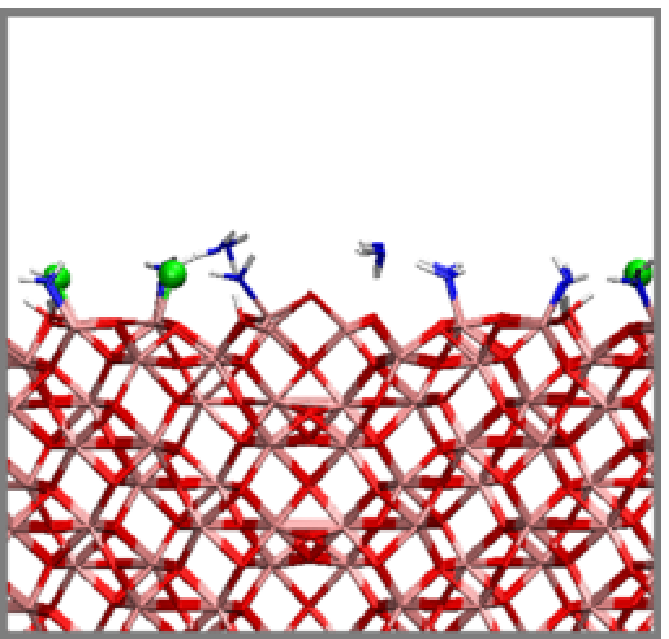} }}
\centerline{\hbox{ (e) \epsfxsize=1.50in \epsfbox{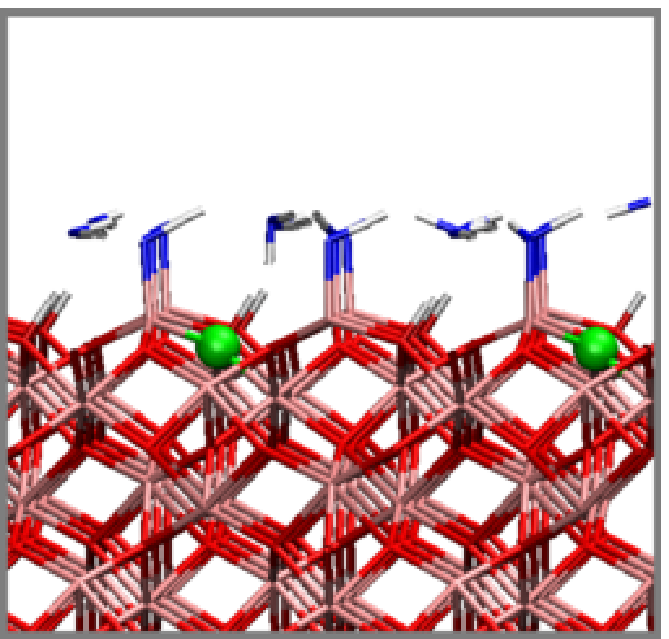}
		   (f) \epsfxsize=1.50in \epsfbox{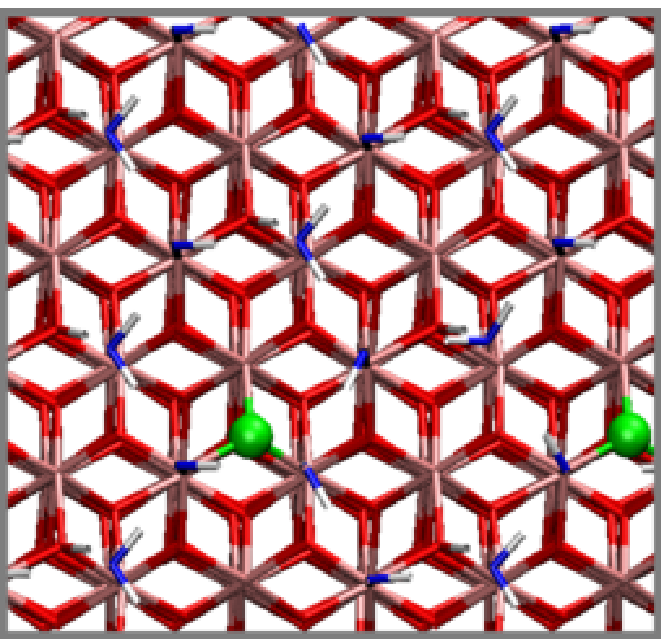} }}
\caption[]
{\label{fig3} \noindent
Detailed atomic structures at deprotonated surfaces.
(a)-(b) $\alpha$-Al$_2$O$_3$ (001) on Al(111) and correspond to
Fig.~\ref{fig2}a;
(c)-(d) $\alpha$-Al$_2$O$_3$ with grain boundary on Al(111) (Fig.~\ref{fig2}b);
(e)-(f) $\alpha$-Al$_2$O$_3$ (001) on Au(111) (Fig.~\ref{fig2}c).
Pink, red, and blue depict Al, O, and O(water) atoms.  Green spheres
depict O atoms from which H$^+$ are removed.
}
\end{figure}

In some cases, surface proton vacancies are introduced by removing H atoms
from H$_2$O on the oxide outer surface; these H are chosen at spatially
separated locations (Fig.~\ref{fig3}).  \color{black} In Fig.~\ref{fig3}a-d,
H$^+$ are removed from H$_2$O molecules; these OH groups newly created
are coordinated to a single Al$^{3+}$.  In Fig.~\ref{fig3}e-f, H$^+$ are removed
from O$^{2-}$, coordinated to three Al$^{3+}$ cations, which are part of the
Al$_2$O$_3$ surface prior to addition of water; hence no new OH group is
created.  These deprotonation schemes are solely meant to mimic
the electrostatic environments arising from negatively-charge salt anions not
present in our simulations; the precise locations of deprotonated sites will
have quantitative effects but qualitatively speaking they are
not considered critical.  \color{black}

V$_{\rm O}$ is introduced by removing one oxygen atom each at various
positions in the crystalline oxide film.  We assume that the atomic
length-scale structural heterogeneity at the metal/oxide interface does not
strongly affect V$_{\rm O}$ properties, and that each O-atom removed is
representative of all V$_{\rm O}$ at that $z$ position.  This is one advantage
of crystalline oxide models; in amorphous Al$_2$O$_3$ films, V$_{\rm O}$
formation energies may not systematically vary with $z$.  The exception is
Fig.~\ref{fig2}b, where V$_{\rm O}$'s along the grain boundary are different
from V$_{\rm O}$ in bulk-like oxide regions; only the former are considered.  

Cl$^-$ are inserted into these V$_{\rm O}^{q+}$; in effect, they substitute
for O in the lattice.  Calculating Cl$^-$ insertion energetics involves the
following steps in a thermodynamic cycle:
\begin{eqnarray}
{\rm Cl}^- {\rm (aq)} &\rightarrow& {\rm Cl}^- {\rm (g)} \nonumber \\
{\rm Cl}^- {\rm (g)} &\rightarrow& {\rm Cl} {\rm (g)} + e^- {\rm (g)} .
	\label{eqcl4}
\end{eqnarray}
Here ``(g)'' denotes the gas phase.  The two steps in Eq.~\ref{eqcl4} yield
the Cl$^-$ dehydration free energy ($\Delta G_{\rm hyd}$=-3.32~eV),\cite{ion}
and the electron affinity (EA) of the Cl atom in vacuum, respectively.  Using
the PBE functional and a 10~\AA$^3$ simulation cell with one Cl$^-$, we predict
that EA for Cl$^-$is 3.42~eV.  It is 0.20~eV below the experimental gas phase
value of 3.62-3.72~eV;\cite{clea} here we adopt a 3.62~eV value.  

To examine Cl$^-$ replacing surface OH$^-$ groups (Eq.~\ref{eqcl2}), we apply
$\alpha$-Al$_2$O$_3$ (0001) slabs with no metal present;
this is adequate because no $e^-$ is transferred alongside these
substitutions (Eq.~\ref{eqcl2}).  The standard state (1.0~M concentration)
OH$^-$ hydration free energy, needed here, is -4.53~eV taken from the
literature.\cite{dixon}
Other simulation cells contain only $\alpha$-Al$_2$O$_3$ in bulk crystal
configurations.  These are used to compare the band structures
and charge-neutral oxygen vacancy orbital energy levels of the oxide computed
using the PBE and the more accurate HSE06 functionals via Fig.~\ref{fig1}a-c.

\color{black}
We apply the Bader charge method\cite{bader}  to examine the change of
charge distribution in the different zones of the slab models arising
from insertion of Cl$^-$ into V$_{\rm O}^{n+}$ or deprotonation on the
hydroxylated oxide surface.  The three relevant zones ``$Z$'' are metal, oxide,
and surface water (which may be hydrolysed, split into H$^+$ and OH$^-$).
Bader charges, like all charge-decomposition schemes, are approximate,
especially given the fact that Al metal ions at the interface may be
assigned partial charges.  To facilitate analysis, we only consider
changes in charge in a zone ($\Delta c(Z)$) relative to a reference.
For each system with surface deprotonation but no Cl$^-$ , the reference is
the same slab but without surface deprotonation.  For each system with Cl$^-$
inserted into a V$_{\rm O}$, the reference is the slab with the same amount
of surface deprotonation but no Cl$^-$ or V$_{\rm O}$.  We also assign
all metal atoms to be in the metal zone, including Al$^{3+}$ inside the
oxide.  This is reasonable because Al only exists as Al$^{3+}$ inside the
oxide and are found to exhibit zero Bader charges.  At the metal/oxide
interface, it is difficult to definitively assign whether a Al atom is in
the metal or the oxide zone; our scheme circumvents this ambiguity.
\color{black}

\begin{figure}
\centerline{\hbox{  \epsfxsize=3.50in \epsfbox{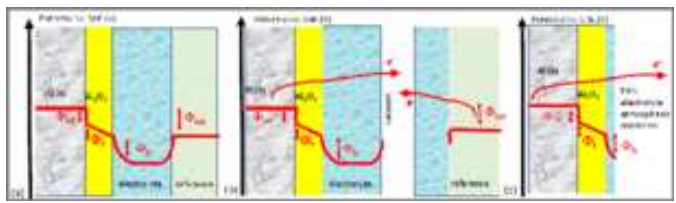} }}
\caption[]
{\label{fig4} \noindent
(a)-(c) Definitions of the $\Phi_{\rm mf}$, $\Phi_{\rm f}$, and $\Phi_{\rm fs}$
contributions to the total potential in full and half cell configurations.
}
\end{figure}

\subsection{Band Alignment and Voltage}
\label{voltage}

In this section, we emphasize that local contributions to the voltage
are more relevant to V$_{\rm O}$ charging than the applied voltage 
($\Phi_{\rm ext}$) itself.  

Experimentally, voltage differences are measured using a reference electrode,
which exhibits a $E_{\rm F}$ different from the working electrode.  The applied
voltage $\Phi_{\rm ext}$ contains several contributions (Fig.~\ref{fig4}a):
\begin{equation}
\Phi_{\rm ext}= \Phi_{\rm mf}+ \Phi_{\rm f}+ \Phi_{\rm fs} , \label{eqphi}
\end{equation}
Here ``m'' denotes metal, ``f'' is the oxide film, ``s'' is the \color{black}
``solution'' \color{black}
or electrolyte, and ``mf'' and ``fs'' are the interfaces.  
Fig.~\ref{fig4}b closely resembles the voltage construction in the point
defect model (Fig.~2 in Ref.~\onlinecite{macdonald81a}).  These
voltage contributions are also discussed in Ref.~\onlinecite{marks}, which
however does not apply DFT calculations with explicit electric fields
across the oxide layer.

DFT is an electronic structure ground state theory and supports only one
$E_{\rm F}$.  The most readily available reference ``electrode'' is vacuum.  We
inject vacuum into the liquid electrolyte region (Fig.~\ref{fig4}b);
this step is formally exact.  Then, in Fig.~\ref{fig4}c, the reference
electrode is replaced by vacuum, and the absolute voltage referenced against
SHE (${\cal V}_e$) replaces $\Phi_{\rm ext}$ in Fig.~\ref{fig4}a-b.  Following
the Trasatti convention,\cite{trasatti} we define
\begin{equation}
{\cal V}_e = W/|e| - 4.44~V, \label{traseq}
\end{equation}
where $|e|$ is the electronic charge, $W$=$-E_{\rm F}$ is the work function,
and $E_{\rm F}$ is referenced to vacuum.  The work function is cumulatively
modified by the electric double layer (EDL) at each interface.

An important question is whether the model system -- which does not include
a liquid electrolyte -- gives a structure/voltage relation that mimics
experimental conditions.  Our simplified electrolyte models consist of
sub-monolayer H$_2$O on the oxide surface optimized at T=0~K.  In the appendix,
it is shown that this approach is unlikely to provide useful comparison with the
electrochemical measured ${\cal V}_e$ or $\Phi_{\rm ext}$, or for that matter
Kelvin Probe Force Microscopy (KPFM) measurements of work functions in vacuum
environments.\cite{kpfm}  The reason is that, on Al$_2$O$_3$ surfaces, the
oxide-solvent term $\Phi_{\rm fs}$ is very sensitive to the surface water
coverage, structure, and presence/absence of hydrolysis events (H$_2$O +
$>$AlO $\rightarrow$ 2 $>$AlOH).  Instead, we will assume a
voltage of -0.50~V vs.~SHE to mimic the onset of pitting,\cite{pittingvoltage}
and ignore the computed ${\cal V}_e$ (Eq.~\ref{traseq}) for the purpose of
evaluating Eq.~\ref{eqcl1}.  Despite this, the calculated ${\cal V}_e$ 
will prove to be a useful measure of convergence with respect to simulation
cell size.  Calculating accurate ${\cal V}_e$'s for oxide-covered metal
surfaces remains a challenge that will require more future work
(Sec.~\ref{outlook}).

The more relevant point about voltage dependence is that ${\cal V}_e$ or
$\Phi_{\rm ext}$ is made up of several interfacial components
(Eq.~\ref{eqphi}), none of which has been separately measured.  The appendix
will show that the contact potential at the Al/Al$_2$O$_3$ metal-film interface
($\Phi_{\rm mf}$) and the voltage drop in the oxide film ($\Phi_{\rm f}$)
most strongly affect the energy levels of orbitals associated with the valence
band edge relative to the V$_{\rm O}$ energy level ($E_{\rm VoVBE}$) or
relative to $E_{\rm F}$ ($E_{\rm FVBE}$).  In contrast, the potential drop at
the oxide/solvent interface ($\Phi_{\rm fs}$), which resides on the side of
the oxide film opposite to the metallic region (Fig.~\ref{fig4}c), does not
significantly affect $E_{\rm VoVBE}$ or $E_{\rm FVBE}$, as long as the
oxide/vacuum interface has no net charge.  The fact that a charge-neutral
liquid electrolyte, not included in our models, supplies only a global shift of
the band structure is taken as justification of our replacement of the
predicted ${\cal V}_e$ with ${\cal V}_e$=-0.5~V vs.~SHE.

A charge-neutral metal surface and a charge-neutral oxide film yields
$\Phi_{\rm f}$=0~V.  As will be shown, this is not sufficient to generate
charged V$_{\rm O}$.  Finite $\Phi_{\rm f}$ is created by
deprotonating one or more H$_2$O at the film/solvent interface, creating a net
negative surface charge density.  Because the DFT simulation cell is
charge-neutral, this should \color{black}
induce \color{black} positive charges at the metal/film interface.
The H$^+$ vacancies change the average dipole surface density $d$ in the
direction perpendicular to the interface.  A positive change in $d$ increases
the voltage according to the equation\cite{pccp}
\begin{equation}
\Delta {\cal V}_e = 4 \pi \Delta d/A , \label{eqdip}
\end{equation}
where $A$ is the lateral surface area.  Eq.~\ref{traseq} can be readily
evaluated in simulation cells with a vacuum region.  Compared to other DFT
formulations in the literature, we control ${\cal V}_e$ using $e^-$ and
atoms,\cite{neurock,mira,sugino1,voltage,vedge,mira2014} not effective
medium approaches extrinsic to standard DFT.\cite{arias,otani,otani12,otani13}  

\subsection{System Size Effects}
\label{size}

A V$_{\rm O}^{2+}$ residing in the electronically insulating oxide film
should also induce a negative surface charge on the metal slab in contact with
it in the charge-neutral simulation cell.  The charge separation leads to
a significant dipole moment along the $z$-direction.  We apply the  standard
dipole correction to remove the electrostatic coupling between periodic images
in the $z$ direction, separated by the vacuum region.\cite{dipole}  However,
the cell is also periodically replicated in the lateral dimensions.  The
lateral images of the dipole moment repel one another.\cite{geojpcl}  These
spurious electrostatic couplings decrease with the lateral cell area $A$.
Hence we systematically increase the lateral cell dimensions with area $A$
while keeping a single V$_{\rm O}^{2+}$ or Cl$^-$ in the cell.

Another manifestation of system size effect is that ${\cal V}_e$ may depend
on $A$.  Benchtop corrosion experiments are often performed at constant
voltage conditions.  In DFT calculations, introducing a single charged Cl$^-$
or V$_{\rm O}^{2+}$ should occur at constant ${\cal V}_e$ in the limit of
infinite $A$ (Eq.~\ref{eqdip}).  At finite $A$, \color{black} however, 
\color{black} ${\cal V}_e$
is modified by the introduction of charged defects. Systematically increasing
$A$ lessens this problem.  The quantum continuum approach\cite{dabo1} or
local charge compensation methods\cite{schultz} may also be applied to
maintain the system at the same ${\cal V}_e$ without system size increase.
These methods, which require extensions to standard DFT codes, will be
considered in the future.


In contrast, a charge-neutral defect like V$_{\rm O}$ is not associated with
a net charge transfer or change in dipole moment $d$.  The resulting band
structure should be insensitive to $A$.  

\section{Results}

\begin{figure}
\centerline{\hbox{ \epsfxsize=4.00in \epsfbox{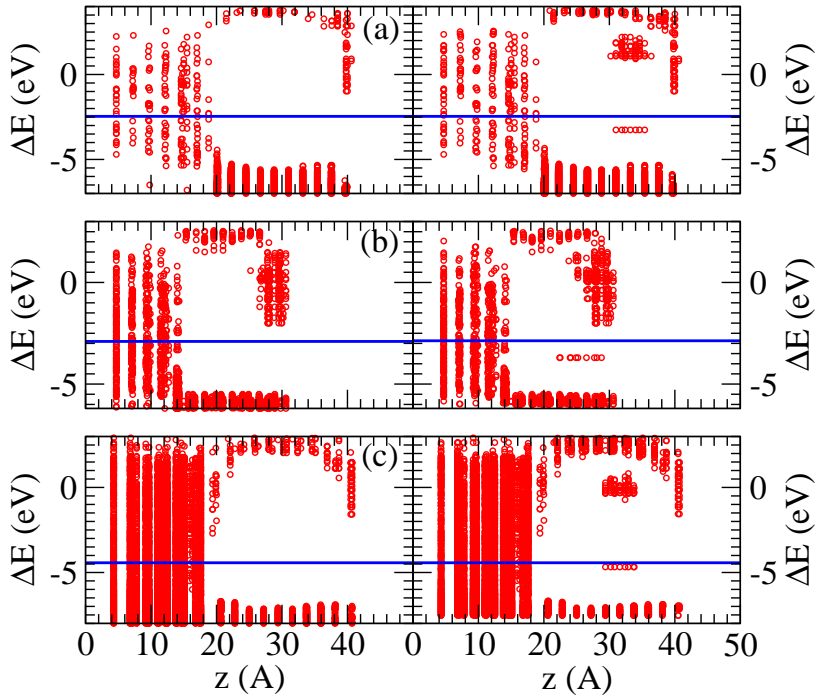} }}
\caption[]
{\label{fig5} \noindent
LDOS of interfacial systems considered in this work.
(a) $\alpha$-Al$_2$O$_3$ (001) on Al(111);
(b) $\alpha$-Al$_2$O$_3$ with grain boundary on Al(111);
(c) $\alpha$-Al$_2$O$_3$ (001) on Au(111).
The left and right panels are with and without V$_{\rm O}$, respectively.
No H$^+$ vacancy exists on any surface in this figure.  Red circles represent
atoms on which a Kohn-Sham orital at a certain energy exceed $>$0.2\%. 
In vacuum regions, no atoms exist and no LDOS is represented therein.
}
\end{figure}

\subsection{Oxgyen Vacancies without Interfaces; Interfaces without Vacancies}

\label{hse06}

Our goal is to study simulation cells with both explicit interfaces and
defects.  But in this section, we first examine the orbital energy level of a
charge-neutral V$_{\rm O}$ in crystalline $\alpha$-Al$_2$O$_3$ relative to
the valence band edge (VBE) ($\Delta E_{\rm VoVBE}$) without Al metal present
(Fig.~\ref{fig1}c), as well as the band offset and the contact potential
between Al(111) and $\alpha$-Al$_2$O$_3$(111) in the absence of V$_{\rm O}$
(Fig.~\ref{fig1}b).  This two-step scheme uses smaller simulation cells, so
that comparison of PBE calculations with those using the more computationally
costly HSE06 DFT functional can be made.  PBE underestimates band gaps and
defect orbital energy levels inside the gap.  HSE06 generally predicts more
accurate orbital alignment, but it is challenging to apply to systems with
800+ atoms like the interfacial cells which can accommodate V$_{\rm O}$
(Table~\ref{table1}).  Subtracting $\Delta E_{\rm VoVBE}$ from the offset
between the Fermi level and the VBE ($\Delta E_{\rm FVBE}$), computed in
two different simulation cells,  gives an approximate alignment between
V$_{\rm O}$ orbitals and the Fermi level.  However, this two-step approximation
is not valid when charge transfer occurs (e.g., if a V$_{\rm O}^{2+}$ is
present), and/or when an electric field exists across the oxide film.

Details of the $\alpha$-Al$_2$O$_3$ cell with one charge neutral V$_{\rm O}$
is  described in Table~\ref{table1}.  The V$_{\rm O}$ is occupied by two
$e^-$ and  is \color{black} an \color{black}
 ``$f$-center.''  The 2$\times$2$\times$2 Brillouin zone
sampling yields 6~special $k$-points.  Near the band gap, the energy level
of each orbital differs by at most 0.04~ eV at different $k$-points.
Hence it is a good approximation to report these values averaged over the 6
$k$-points, instead of computing the entire band sructure of this supercell and
searching for the absolute VBE.  We find that $\Delta E_{\rm VoVBE}$=2.34~eV and
3.12~eV for PBE and HSE06, respectively.  These values are in good agreement
with the
literature.\cite{al2o3_ovac0,al2o3_ovac1,al2o3_ovac2,al2o3_ovac3,al2o3_ovac4}
Increasing the cell size from Al$_{48}$O$_{72}$ to Al$_{108}$O$_{162}$ yields
similar PBE results.

Regarding $\Delta E_{\rm FVBE}$ at explicit Al(111)/$\alpha$-Al$_2$O$_3$(0001)
interfaces, the PBE and HSE06 values are 2.94~eV and 4.08~eV, respectively.
As expected, HSE06 increases the band gap and shifts the valence band to
lower energy levels compared to PBE.  Combining these results and
$\Delta E_{\rm VoVBE}$, the two functionals predict that V$_{\rm O}$ sits
0.60~eV and 0.96~eV below $E_{\rm F}$, respectively (Fig.~\ref{fig1}d-f).
Switching to HSE06 thus lowers the occupied V$_{\rm O}$ orbital level by
-0.36~eV relative to the Fermi level. It would take an extra +0.36~V
applied potential to shift this orbital level, computed using PBE, to a
regime where it will begin to be oxidized (i.e., lose $e^-$ and acquire
positive charge(s)) if we were to use the more accurate HSE06 functional.
Since one of our main theses is that it is more difficult to create
positively-charged V$_{\rm O}$ than is assumed in the literature, switching
to the more accurate HSE06 functional will only reinforce this conclusion.
In the remainder of this paper, we will apply the PBE functional, for which
$\Delta E_{\rm VoVBE}$$<$2.34~eV is a zeroth-order guide for the orbital level
of uncharged V$_{\rm O}$.

$\Delta E_{\rm FVBE}$ contains a contribution from the
metal/oxide contact potential ($\Phi_{mf}$, Fig.~\ref{fig4}a-c).\cite{marks}
$\Phi_{mf}$ can be computed by subtracting the work function in presence of
an oxide film from the work function in the absence of oxide coating the
metal surface, as long as the oxide slab does not exhibit a net dipole moment.
PBE and HSE06 yield -0.50~V and -0.68~V, respectively.  The HSE06 value is
larger by 36\%; this ratio is larger than that between HSE06 and PBE values
at the Li metal/LiF (001) interface.\cite{pccp}

\subsection{Oxgyen Vacancies are Charge Neutral when $\Phi_{\rm f}$=0
\color{black} (no surface deprotonation) \color{black}}
\label{nofield}

Next we consider other metal/oxide interfaces in larger simulation cells.
The panels on the left hand side of Fig.~\ref{fig5} depict the local electronic
densities of state (LDOS) in interfacial supercells without a V$_{\rm O}$.
The valence band edges are flat across the oxide film region, corresponding
to a no-electric field ($\Phi_{\rm f}$=0, Fig.~\ref{fig4}c), ``flat band''
condition.\cite{dabo1}  $\Delta E_{\rm FVBE}$ are 2.94~eV, 1.52~eV, 
and 2.49~eV, respectively, in Fig.~\ref{fig5}a-c.  
The $\alpha$-Al$_2$O$_3$ cases (a) and (c)
are higher than $\Delta E_{\rm VoVBE}$ 2.34~eV, and are consistent with
V$_{\rm O}$ residing below the Fermi level in DFT/PBE calculations.  
We have not calculated $\Delta E_{\rm VoVBE}$ for (b)
because, unlike in oxide films with no defects, this quantity should depend
on the spatial position of the V$_{\rm O}$ and is not single-valued.

\begin{figure}
\centerline{\hbox{ \epsfxsize=4.50in \epsfbox{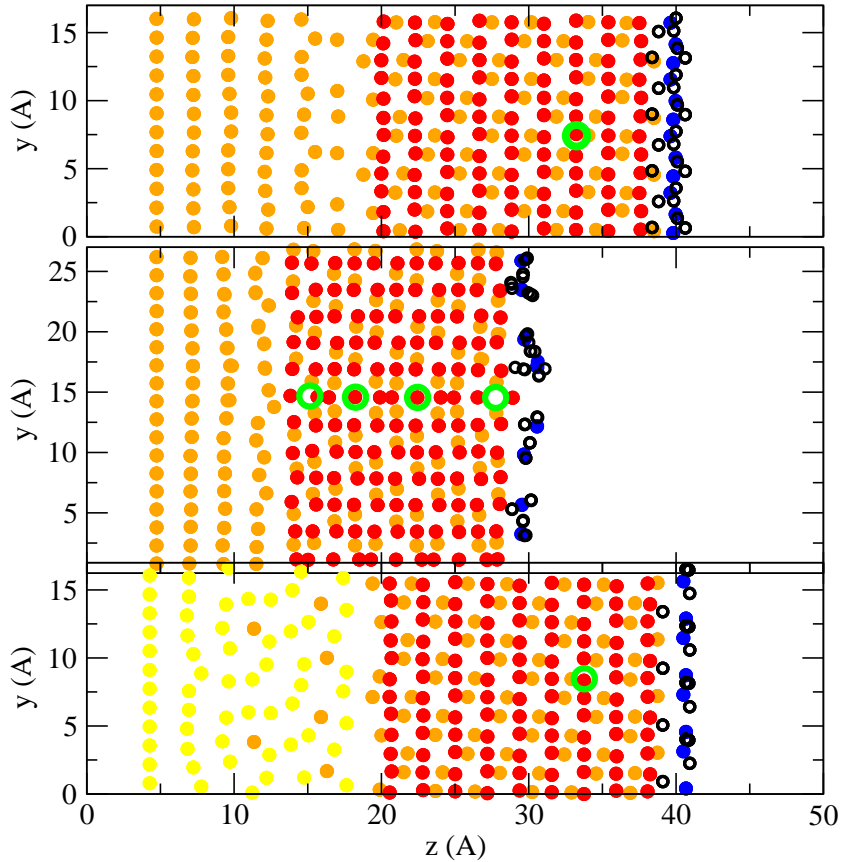} }}
\caption[]
{\label{fig6} \noindent
Illustrations of the positions of the Cl$^-$ and V$_{\rm O}$ discussed in
the text.  The three panels correspond to Fig.~\ref{fig2}a-c, respectively.
\color{black}
Orange, red, blue, yellow, hollow black, and hollow green circles refer
to Al, O, O(water), Au, H, and Cl atoms, respectively. \color{black}
}
\end{figure}

The right hand side panels represent configurations with an explicit
V$_{\rm O}$ each in the oxide film.  Their locations are illustrated in
Fig.~\ref{fig6}.   In all cases, the orbitals associated with the V$_{\rm O}$
lie below $E_{\rm F}$.  This is consistent with the $E_{\rm VoVBE}$
plus $E_{\rm FoVBE}$ estimate applied to the left hand side panels without
V$_{\rm O}$ in the cells.  The
V$_{\rm O}$ orbital level associated with Au(111) (Fig.~\ref{fig5}c) is
closer to $E_{\rm F}$ than on Al(111) (Fig.~\ref{fig5}a), consistent
with the fact that Au is more electropositive and exhibits less tendency to
emit electrons.  These findings constitute the main result of this paper.

The above conclusion that V$_{\rm O}$ lies below $E_{\rm F}$ does not 
qualitatively depend on the spatial location of the V$_{\rm O}$.  In the case
of Fig.~\ref{fig5}b, moving the V$_{\rm O}$ within the grain boundary from
$z$=26.5~\AA\, to 22.4~\AA, 18.3~\AA, or 16.4~\AA\, ($\Delta z$=12.4, 8.3,
4.2, or 2.3~\AA, Fig.~\ref{fig5}) changes the offset between E$_{\rm F}$ and the
V$_{\rm O}$ orbital level from -0.80~eV to -0.27~eV, -0.44~eV, and -0.55~eV.
Here $\Delta z$ is the vertical distance of the V$_{\rm O}$ from the metal/oxide
interface, which is defined as the $z$-position of the lowest lying plane
containing oxygen atoms.  The total energies of these systems are increased
relative to that of the V$_{\rm o}$ at the outermost $z$=26.5~\AA\, position
by 0.31~eV, 0.30, or 0.22~eV respectively; the small energy changes are likely
due to non-constant strain effects at different
film depths.  For the V$_{\rm O}$ at $z$=22.4~\AA\, ($\Delta z$=12.4~\AA), we
also apply spin-polarized DFT to explore the possibility of a spin-polarized
V$_{\rm O}^+$ vacancy.  No charged V$_{\rm O}$ is observed.  This 
spin-polarization issue may be reconsidered in the future using hybrid DFT.

As discussed in Sec.~\ref{dft}, the atomic length-scale structure at the
metal/oxide interface is not known from experiments.  This may give rise to
uncertainties in DFT-predicted $\Phi_{\rm mf}$ on the order of
$<$0.5~V.\cite{wang,bredas}  However, even a +0.5~V shift in $\Phi_{mf}$
does not change the qualitative DFT/PBE conclusion about V$_{\rm O}$ if we 
include the 0.36~eV HSE06 correction (Sec.~\ref{hse06}).
We find that increasing the lateral system size by doubling the $x$- or
$y$-dimension by a factor of two yields no qualitative and minimal
quantitative changes in  these conclusions about uncharged V$_{\rm O}$,
as anticipated in Sec.~\ref{size}.\cite{cr_interface}

Finally, we discuss ${\cal V}_e$ predictions.
The voltages ${\cal V}_e$ associated with the left side of Fig.~\ref{fig5}a-c
are -1.98~V, -1.54~V, and -0.01~V, respectively.  The
corresponding ${\cal V}_e$ on the right hand side of these panels, where 
V$_{\rm O}$ exist, are almost unchanged.  However, the appendix shows that
such ${\cal V}_e$ are very sensitive to the H$_2$O adsorption configuration, 
especially when hydrolysis event occurs, leading to global shifts in the
electronic band structure due to the change in $\Phi_{\rm fs}$
(Fig.~\ref{fig4}).  This is unlike the case of \color{black} organic 
\color{black} solvent electrolytes
used in batteries.\cite{pccp} One reason is that each water hydrolysis event
causes significant charge separation.  Hence these ${\cal V}_e$ cannot be
considered definitive, nor should they be compared with experiments.  

\subsection{Charged Oxgyen Vacancy Exists With Field in Oxide Film
\color{black} (with surface deprotonation) \color{black}}
\label{field}

Next we show that a positive electric field pointing from metal to outer oxide
film surface can create doubly charged oxygen vacancies.  
Electric fields within the oxide can be caused by an enhanced concentration of
cations or anions at the liquid-solid interface.  Since ions and liquid
electrolytes are excluded from our calculations, we instead create 2-3
H-vacancies among AlOH groups on the outer surface, forming AlO$^-$
(Sec.~\ref{dft}).  These AlO$^-$ groups \color{black} should \color{black}
induce a compensating positive
charge on the metal/oxide interface, a positive electric field, and an increased
${\cal V}_e$ via Eq.~\ref{eqdip}.  If a charged V$_{\rm O}^{2+}$ is present, 
it should provide a partial balancing charge for the AlO$^-$ surface groups.

\begin{figure}
\centerline{\hbox{ \epsfxsize=4.00in \epsfbox{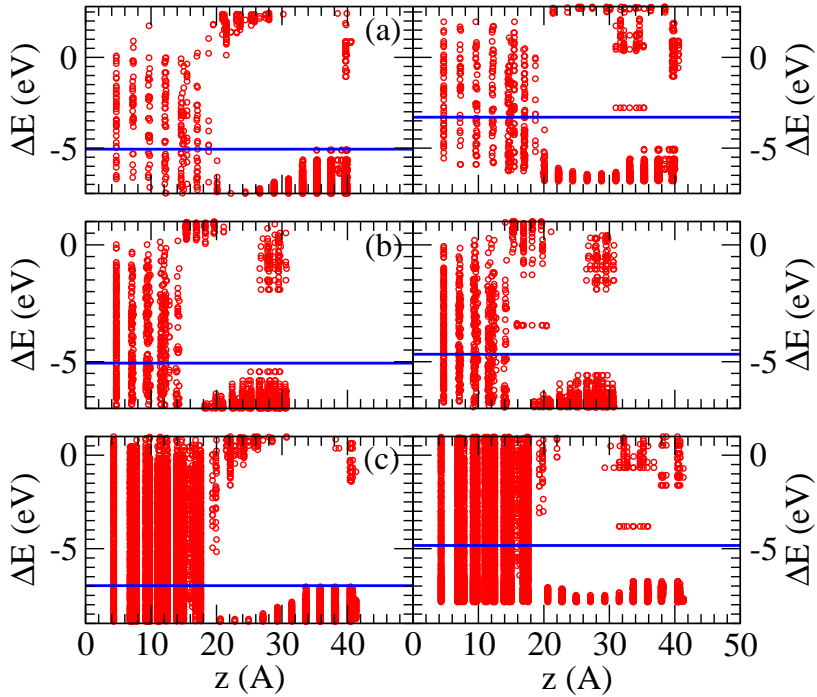} }}
\caption[]
{\label{fig7} \noindent
LDOS of interfacial systems considered in this work.
(a) $\alpha$-Al$_2$O$_3$ (001) on Al(111);
(b) $\alpha$-Al$_2$O$_3$ with grain boundary on Al(111);
(c) $\alpha$-Al$_2$O$_3$ (001) on Au(111).
Red circles denote atoms with Kohm-Sham orbital contributions exceeding 
0.2\%.  The right hand column panels contain a V$_{\rm O}$ while the left hand
columns do not.  Unlike Fig.~\ref{fig5}, there are 2-3 H$^+$ vacancies
on the outer surfaces of the oxides.
}
\end{figure}

Fig.~\ref{fig7}a-c correspond to Fig.~\ref{fig5}a-c, except that the LDOS
for an O-vacancy closer to the metal surface than the one in Fig.~\ref{fig5}b
is depicted in Fig.~\ref{fig7}b (see below).  The left hand column
depicts LDOS with a field but without V$_{\rm O}$.  The valence band edges
show that potential gradients indeed exist.  Note that the field
directions and the slopes associated with  $\Phi_{\rm f}$ are opposite to
those depicted in Fig.~\ref{fig1}a-c or in classic PDM
schematics.\cite{macdonald81a,marks}  The computed ${\cal V}_e$ of these
systems are 0.96~V, 0.62~V, and 2.54~V, respectively.  They 
are much higher than those associated with Fig.~\ref{fig5}; the changes
are $\Delta {\cal V}_e$=+2.94~V, +2.16~V, and +2.55~V.
Experimentally, raising the voltage is consistent with increased pitting
corrosion.  As discussed in Sec.~\ref{voltage}, we do not claim that our
computed ${\cal V}_e$ are quantitative.

\color{black}
Table~\ref{table2} depicts the changes in aggregate Bader charges
($\Delta c(Z)$) in the zones $Z$=metal, oxide, and water due to surface
deprotonation in systems without oxygen vacancies.
With the $\alpha$-Al$_2$O$_3$ slabs on Al(111) and
Au(111) metal surfaces (Fig.~\ref{fig2}a, Fig.~\ref{fig2}c), there are
approximately +2$|e|$ charges induced in the metal zone, where $|e|$
is the electronic charge.  This is reasonable because the two H-atoms
removed on the surface are expected to be H$^+$, inducing equal but
opposite charge in the metal zone.  The demarcation of charges in
the oxide and water zones appears more ambiguous, partly because of 
hydrolysis reactions arguably turn some O$^{2-}$ on the oxide surface
into OH, but we have not reassigned atoms across zones.  The grain boundary
system (Fig.~\ref{fig2}b) has three surface OH groups deprotonated, while
the Bader analysis suggests that a +4$|e|$ is induced on the metal zone.
The ambiguity likely arises from assigning all metal atoms at
the metal/oxide interface to the metal zone.

\begin{table}\centering
\begin{tabular}{l|c|r|r|r} \hline
figure  & deprot. & metal & oxide & water \\ \hline
Fig.~\ref{fig2}a & 2 & +1.52 & -0.28 & -1.26 \\
Fig.~\ref{fig2}b & 3 & +3.95 & -1.43 & -2.52 \\
Fig.~\ref{fig2}c & 2 & +1.65 &  0.56 & -2.74 \\ \hline
\end{tabular}
\caption[]
{\label{table2} \noindent
Change in charge in deprotonated slabs in each zone $(\Delta c(Z)$ in units of
$|e|$), counting both electrons and nuclei, relative to the slab
without deprotonation.  No V$_{\rm O}$ or Cl$^-$ exists in these slabs.
The number of H atoms removed from the surface is ``deprot.''
}
\end{table}
\color{black}

The right hand columns of Fig.~\ref{fig7} depict LDOS with an explicit
V$_{\rm O}$ in each simulation cell.  In all cases, the electric field has
moved the orbitals associated with the defect into the gap, creating a
V$_{\rm O}^{2+}$.  We also perform spin-polarized DFT
calculations in Fig.~\ref{fig7}b to check if singly charged V$_{\rm O}^+$
occurs.  The result is the same as in the non-spin-polarized DFT calculation.
Because of the potential gradient, V$_{\rm O}^{2+}$ more readily occurs in
the outer regions of the oxide film.  Moving the V$_{\rm O}$ to $z$=26.7~\AA\,
($\Delta z$=12.6~\AA) from where it is ($z$=18.2~\AA,
$\Delta z$=4.1~\AA\, in Fig.~\ref{fig7}b) raises the V$_{\rm O}$ vacancy to
1.98~eV above $E_{\rm F}$ instead of 1.24~eV above $E_{\rm F}$.  The orbital
level associated with the V$_{\rm O}$ at $z$=26.7~\AA\, for the system with
Au(111) almost merges with the conduction band, and is not shown; instead
Fig.~\ref{fig7}c depicts the LDOS of a V$_{\rm O}$ closer to the metal
surface, at $z$=18.2~\AA.

As mentioned in Sec.~\ref{size}, simulation size effects is more important
when charged defects exist.  Doubling one of the lateral dimensions of the
simulation cell associated with Fig.~\ref{fig7}b is found to yield a charged
V$_{\rm O}$ orbital level at 1.27~eV above $E_{\rm F}$.  This
$\Delta E_{\rm FVBE}$ is only 0.05~eV higher than that for the smaller cell
size.  However, for Fig.~\ref{fig7}c, doubling the system size increases
$\Delta E_{\rm FVBE}$ from 1.00~eV to 1.85~eV.  These variations are
not straight-forward to analyze because there are charges on the
oxide outer surface, in the V$_{\rm O}^{2+}$, and at the metal-film interface.
Regarding Fig.~\ref{fig7}b, moving the V$_{\rm O}^{2+}$ from $z$=26.7~\AA\,
($\Delta z$=12.6~\AA)  to $z$=18.2~\AA\, ($\Delta z$=4.1~\AA) leads to
-4.18~eV and -4.59~eV changes in the total energy depending on whether
we use the base simulation cell (Table~\ref{table1})
or a cell doubled in the lateral dimension.  In all these cases, increasing
the cell size does not lead to the V$_{\rm O}^{2+}$ becoming charge neutral.
Thus system size has quantitative but no qualitative effects.
We have not considered Al$^{3+}$ cation motion.  However, such cations
should also be attracted towards the negatively charged oxide surface and
promote anodic corrosion when a high voltage is applied.


We conclude that V$_{\rm O}^{2+}$ most likely requires a positive electric
field and a higher effective voltage than that under flat band conditions to
exist.  This view about Al$_2$O$_3$ films on Al metal surfaces
is predicated on the assumption that $e^-$ can tunnel through the oxide
and attain the electronic ground state.  If the oxide is sufficiently
thick, non-equilibrium electronic configurations may persist.

\subsection{Cl$^-$ Electrochemical insertion into Oxygen Vacancies}
\label{cle}

\color{black}
Next we consider the energetics of Cl$^-$ insertion into pre-existing
V$_{\rm O}$, charged or otherwise (Eq.~\ref{eqcl1}).  Unlike Sec.~\ref{nofield}
and~\ref{field}, simulation cells both with and without surface charges are
considered.  \color{black}  We assume that oxygen
vacancies in the oxide film are not mobile within the time scale of Cl
insertion.  Cl$^-$ insertion kinetics are not considered in this work.
The overall change in the charge-neutral simulation cell is
\begin{equation}
({\rm Al}^{n-}|{\rm oxide}) + {\rm V}_{\rm O}^{n+} + {\rm Cl}^- (aq) \rightarrow
({\rm Al}^{(m)-}{\rm |oxide}) + {\rm Cl}_{\rm Vo}^{m+} + e^- .
				\label{eqcl3}
\end{equation}
The $e^-$ released on the right side goes to $E_{\rm F}$ in the Al metal
region, effectively ``outside'' the simulation cell, yielding an energy gain
that is equal to $|e|$$E_{\rm F}$.  As discussed in Sec.~\ref{voltage}, we
impose ${\cal V}_e$=-0.50~V vs.~SHE, which corresponds to $E_{\rm F}$=-3.94~eV,
instead of using the ${\cal V}_e$ and $E_{\rm F}$ computed for each slab.

First we consider no surface deprotonation, in which case the V$_{\rm O}$ 
are uncharged hosts.  With the $E_{\rm F}$ and $\Delta G_{\rm hyd}$ discussed
above, and other input listed in Sec.~\ref{dft}, the energies of inserting a
Cl$^-$ into the V$_{\rm O}$ associated with Fig.~\ref{fig5}a-c are 
$\Delta E_{\rm Cl-O}$=+1.14~eV, +0.54~eV, and +0.61~eV at -0.50~V vs.~SHE.
Appling a voltage $>$-0.5~V (higher work function) would further favor
Eq.~\ref{eqcl3}.  Cl$^-$ is a larger anion than oxygen, and inserting Cl$^-$
into V$_{\rm O}$ in the dense $\alpha$-Al$_2$O$_3$ proves less energetically
favorable than into V$_{\rm O}$ in the grain boundary (Fig.~\ref{fig5}b), where
there is more space.  Focusing on the grain boundary in $\alpha$-Al$_2$O$_3$,
inserting Cl$^-$ into V$_{\rm O}$ at distances progressively closer to the
oxide outer surface $z$=15.7~\AA, 18.2~\AA, 22.4~\AA, and 26.5~\AA\, 
($\Delta z$=1.6~\AA, 4.1~\AA, 8.3~\AA, and 12.4~\AA\, respectively) yields
$\Delta E_{\rm Cl-O}$=-0.21~eV, -0.24~eV, -0.22~eV, and +0.54~eV.
These energies are similar except when Cl$^-$ is near the oxide surface.  

These $\Delta E_{\rm Cl-O}$ \color{black} values \color{black}
are computed using the baseline simulation cells
(Table~\ref{table1}).  Table~\ref{table3} shows that both $\Delta E_{\rm Cl-O}$
and the calculated voltage ${\cal V}_e$ can exhibit some dependence on the
lateral surface area of the simulation cell, as anticipated in Sec.~\ref{size}.
The convergence rates are not uniform and depend on materials and location
of the Cl$^-$.  However, in the 2$\times$2 supercells where the lateral cell
sizes exceed 30~\AA, $\Delta E_{\rm Cl-O}$ converge
to within 0.06~eV, while $\Delta {\cal V}_e$, which measure the difference
in ${\cal V}_e$ with and without the Cl$^-$, trend towards zero as they
should.  This suggests that the largest simulation cells considered in
Table~\ref{table3}, containing 3000-4000 atoms, are sufficient for computing
Cl$^-$ insertion energies in most cases.  More significantly, Cl$^-$ insertion
into the grain boundary (Fig.~\ref{fig5}b) is found to be energetically
favorable at an applied -0.50~V voltage, as long as it is sufficiently
far from the surface.  Special care is taken to ensure that, as the simulation
cells are expanded in the lateral directions, the water configurations are
unchanged.

\begin{table}\centering
\begin{tabular}{l|l|r|r|r} \hline
system &  property & 1$\times$1 & 1$\times$2 & 2$\times$2 \\ \hline
Fig.~\ref{fig2}a & $\Delta E_{\rm Cl-O}$ & +1.14 eV & +0.93 eV & +0.89 eV \\
& $\Delta {\cal V}_e$ & -0.88 V & -0.50 V & -0.23 V \\
Fig.~\ref{fig2}b & $\Delta E_{\rm Cl-O}$ & -0.24 eV & -0.27 eV & -0.22 eV \\
(*) & $\Delta {\cal V}_e$ & -0.20 V & -0.08 V & -0.05 V \\
Fig.~\ref{fig2}c & $\Delta E_{\rm Cl-O}$ & +0.61 eV & +0.07 eV & +0.01 eV \\
& $\Delta {\cal V}_e$ & -0.73 V & -0.33 V & -0.11 V \\
\hline
\end{tabular}
\caption[]
{\label{table3} \noindent
Energy cost ($\Delta E_{\rm Cl-O}$) of Cl$^-$ insertion into uncharged
V$_{\rm O}$ as a function of simulation cell size.  $\Delta {\cal V}_e$ is
the computed voltage with the inserted Cl$^-$ versus without; it is zero at
infinite system size and is another measure of convergence.   $n$$\times$$m$
indicates supercell sizes as multiples of each baseline simulation cell.
(*) The V$_{\rm O}$ considered is at $z$=18.2~\AA, not $z$=26.5~\AA\, depicted
in Fig.~\ref{fig5}.
}
\end{table}

Next we consider Cl$^-$ insertion into V$_{\rm O}^{2+}$ in two selected cases.
Adding a Cl$^-$ to the V$_{\rm O}^{2+}$ at $z$=18.2~\AA\, ($\Delta z$=4.1~\AA)
associated with Fig.~\ref{fig7}b yields a -1.94~eV exothermicity, compared with
-0.24~eV without deprotonation at the same lateral cell size.  
These results are in qualitative agreement with the perspective that higher
applied voltages lead to more ready Cl$^-$ insertion into passivating oxide
films.\cite{natishan2000}  However, if the Cl$^-$ were inserted into a
V$_{\rm O}^{2+}$ at $z$=26.5~\AA\, ($\Delta z$=12.4~\AA), $\Delta E_{\rm Cl-O}$
is increased from -0.22~eV without deprotonation to +0.84~eV with protonation.
This increase in $\Delta E$ is likely the result of strong Coulomb repulsion
between the Cl$^-$ and the AlO$^-$ groups at the surface.  Since larger
simulation cells should make $\Delta E_{\rm Cl-O}$ even more favorable
(Sec.~\ref{size}), we have not considered larger cells once 
$\Delta E_{\rm Cl-O}<0$; the qualitative conclusion that Cl$^-$ insertion
into V$_{\rm O}^{2+}$ is in general favorable would be unchanged.

LDOS plots for systems with Cl$^-$ insertion (not shown) do not reveal states
in the gap in the oxide region for either V$_{\rm O}$ or V$_{\rm O}^{2+}$,
suggesting that the Cl$^-$ has driven the $f$-center orbital(s) associated
with the V$_{\rm O}$ into the valence or conduction band.  
\color{black}
Bader charge analysis for Cl$^-$ insertion into V$_{\rm O}$ yields expected
results.  Each Cl$^-$ is found to carry $\sim$8 $e^-$.  So Cl$^-$ is indeed
a monovalent anion in the lattice, giving it a net $\sim$$+|e|$ change in
charge when substituting for an O$^{2-}$ in the lattice.  This induces
a $\sim$-$|e|$ net charge in the metal zone regardless of whether there is
deprotonation in the system (Table~\ref{table4}).

\begin{table}\centering
\begin{tabular}{l|c|r|r|r} \hline
figure  & deprot. & metal & oxide & water \\ \hline
Fig.~\ref{fig2}a & 0 & -0.93 &  0.92 &  0.01 \\
Fig.~\ref{fig2}a & 2 & -0.75 &  0.88 &  0.12 \\
Fig.~\ref{fig2}b & 0 & -0.99 &  0.99 &  0.00 \\
Fig.~\ref{fig2}b & 3 & -1.10 &  1.11 & -0.01 \\
Fig.~\ref{fig2}c & 0 & -0.85 &  0.84 &  0.01 \\ \hline
Fig.~\ref{fig2}c & 2 & -0.86 &  0.84 &  0.01 \\ \hline
\end{tabular}
\caption[]
{\label{table4} \noindent
Change in charge (in units of $|e|$) due to Cl$^-$ substitution for an
O$^{2-}$ in each zone $(\Delta c(Z))$, counting both electrons and nuclei,
relative to the slab without a Cl$^-$; the number of H atoms removed
from the surface is ``deprot.''  Unlike in Table~\ref{table2}, the number
of surface deprotonation is the same for the reference and the target slab.
}
\end{table}
\color{black}

\subsection{Cl$^-$ $\rightarrow$ OH$^-$ Substitution}

\begin{figure}
\centerline{\hbox{ \epsfxsize=2.00in \epsfbox{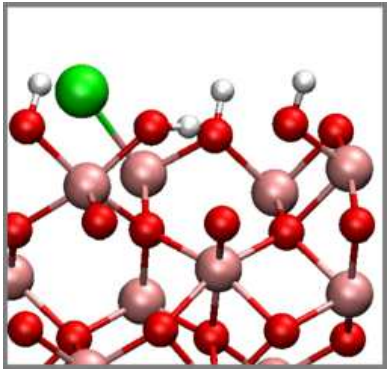} }}
\caption[]
{\label{fig8} \noindent
Substituting a Cl$^-$ for a surface OH$^-$ group on the
$\alpha$-Al$_2$O$_3$ (0001) surface.
}
\end{figure}

Finally, for completeness, we consider Eq.~\ref{eqcl2} using the same
Cl$^-$ $\Delta G_{\rm hyd}$ we have applied in this work.  Eq.~\ref{eqcl2}
is not an electrochemical process; no electron is exchanged with the metal.
The $\alpha$-Al$_2$O$_3$ (0001) reactant and product surface slabs used in
the calculations have stoichiometries Al$_{68}$O$_{54}$H$_6$ and
Al$_{68}$O$_{54}$H$_5$Cl (Fig.~\ref{fig8}), respectively.  
Eq.~\ref{eqcl2} is found to be endothermic by $\Delta E$=+1.54~eV at
standard state for both Cl$^-$ and OH$^-$ anions.  

``Standard state'' means that [OH$^-$]=1.0~M, which is pH=14.  $\Delta E$
at lower pH can be estimated using the Nernst equation where each
decade of OH$^-$ product dilution reduces the endothermicity by 0.0592~eV.
At pH=7 or pH=0, $\Delta E$ is lowered to 1.12~eV and 0.71~eV, respectively.
However, they remain energetically unfavorable.  These findings  dovetail with
the geochemistry expectation that Cl$^-$ does not react with surface AlOH
groups at Al$_2$O$_3$/water interfaces.  This prediction is qualitatively
similar to that in Ref.~\onlinecite{cl-al2o3} which applies a different
DFT functional. Ref.~\onlinecite{cl-al2o3} predicts that Cl substitution on the
Al$_2$O$_3$ (100) surface is favorable; we have not re-examined this 
possibility.  DFT calculations have been applied to examine Cl$^-$ substitution
or insertion on other oxide 
surfaces,\cite{cl-nio,cl-cr2o3a,cl-cr2o3b,cl-fe2o3a,cl-fe2o3b} but the
Cl$^-$ reference states are generally different from that applied in this work.
Potential-of-mean-force simulations for desorption of Cl$^-$ from the surface
in liquid water conditions may be applied to to re-examine the desorption free
energy in the future.\cite{geojpcl}

\section{Outlook}
\label{outlook}

We emphasize the need to use DFT simulation cells with explicit
metal/passivating oxide interfaces.  Even though such calculations are costly,
approximations are involved, and currently insufficient experimental input is
available to pinpoint all structural/electronic interfacial properties, they
can yield qualitatively new insights regarding electric fields,
contact potentials, band-offsets, and charge-transfer effects which are not
available in calculations without explicit metal/oxide interfaces.  Our
modeling work provides a renewed impetus for experimental characterization of
metal/passivating oxide interfaces.

Computational voltage control, and correlation of predicted voltages with the
onset associated with pitting, remain challenging.  This is because the
potential drop inside the oxide film ($\Phi_{\rm f}$), the contact potential at
the metal/oxide film interface ($\Phi_{\rm mf}$), and the oxide/electrolyte
interfacial contribution ($\Phi_{\rm fs}$) all add up to the single-valued
applied voltage ($\Phi_{\rm ext}$).  An overall charge-neutral
oxide/electrolyte interface is predicted to yield only a global shift 
($\Phi_{\rm fs}$) to the electronic band alignment, while changes in
the metal/film interface yield large scale changes in the band structure.
Future experimental work that isolates these separate contributions would
be invaluable.  Alternatively, the elucidation of the definitive atomic
length-scale structure at the metal/oxide interface via cross-sectional
imaging technique will enable more accurate DFT predictions of
these quantities.  DFT functional accuracy, simulation cell sizes, atomic
length-scale interfacial structures, surface charges, oxide film thickness,
and oxide phase specificity (including the possibility of amorphous oxides) all
influence the predictions.  Electric double layers in the aqueous electrolyte
outside the oxide surface need to be explicitly included (say, via classical
force field molecular dynamics simulations, or via the quantum continuum
approach\cite{campbell1}), and added to the DFT work functions to afford 
quantitative comparisons with experiment voltages.  The roles of salt and
carbon dioxide on atmospheric corrosion should also to addressed in these
models.\cite{schaller}  Despite these challenges, our work paves the way for
further computational studies of the complex but crucial passivating oxide
film interfaces relevant to corrosion phenomena.

\section{conclusion}
\label{conclusion}

\color{black}
In this work, 
we examine the charges in oxygen vacancies (V$_{\rm O}$) inside crystalline
Al$_2$O$_3$ model surface films covering Al and Au (111) surfaces.  Several
models of metal/oxide interfaces, with and without grain boundaries, are
considered.   In all cases, when there is no electric field in the oxide film
(``flat-band'' configuration), we predict that V$_{\rm O}$ is charge-neutral.
Creating negative surface charges by deprotonation of surface OH groups
generates electric fields.  These fields are needed to yield V$_{\rm O}^{2+}$
which are postulated to be crucial in the point defect model (PDM) widely used
to analyze corrosion behavior.\cite{macdonald81a}  The required potential
gradients have slopes opposite to those typically seen in corrosion study
schematics.\cite{macdonald81a}  V$_{\rm O}^{2+}$ preferentially resides near
the negatively charged oxide outer surfaces, while charge neutral
V$_{\rm O}$ does not.  Future DFT work along these lines for metals
like Fe and Ni, which are more electropositive than Al and should support
V$_{\rm O}^{2+}$ more readily, may be of significant interest, although
Fe(III) ions in Fe$_2$O$_3$ oxide films may undergo redox changes and
complicate the analysis.

Cl$^-$ is predicted to be energetically favorable when inserted into some
V$_{\rm O}$ or V$_{\rm O}^{2+}$, depending on the oxide strucure, existence
of grain boundary, and applied voltage.  In contrast, Cl$^-$ substitution of
OH$^-$ groups coordinated to surface Al cations does not depend on voltage
and is predicted to be energetically unfavorable on flat $\alpha$-Al$_2$O$_3$
(0001) surfaces.  Such substitutions have been inferred in analysis of EXAFS
studies;\cite{natishan2017} they may occur on other crystal
facets,\cite{cl-al2o3} in oxide pores, or on amorphous oxide surfaces.
\color{black}

\section*{Appendix: Structure/voltage relation at different interfaces}

In this appendix, we seek to isolate the effects of the metal-film interface
and of the film-solvent interface using a combinatorial approach.  We focus
on the Al (111)/$\alpha$-Al$_2$O$_3$ (0001) system.  No oxide-surface
H$^+$ vacancy, V$_{\rm O}$, or Cl$^-$ insertion is considered in this
appendix, which allows us to use smaller simulation cells.  A surface unit
cell of lateral dimension 4.81$\times$8.33~\AA$^2$ is applied in all cases.
This is a ``doubled'' surface cell in the sense that it has twice the lateral
area of the normal $\alpha$-Al$_2$O$_3$ (0001) surface cell.

\begin{figure}
\centerline{\hbox{ (a) \epsfxsize=2.05in \epsfbox{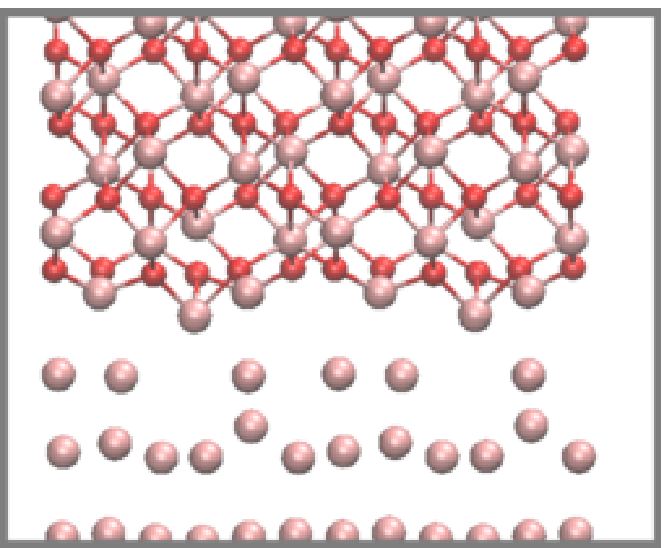} 
		   \epsfxsize=1.95in \epsfbox{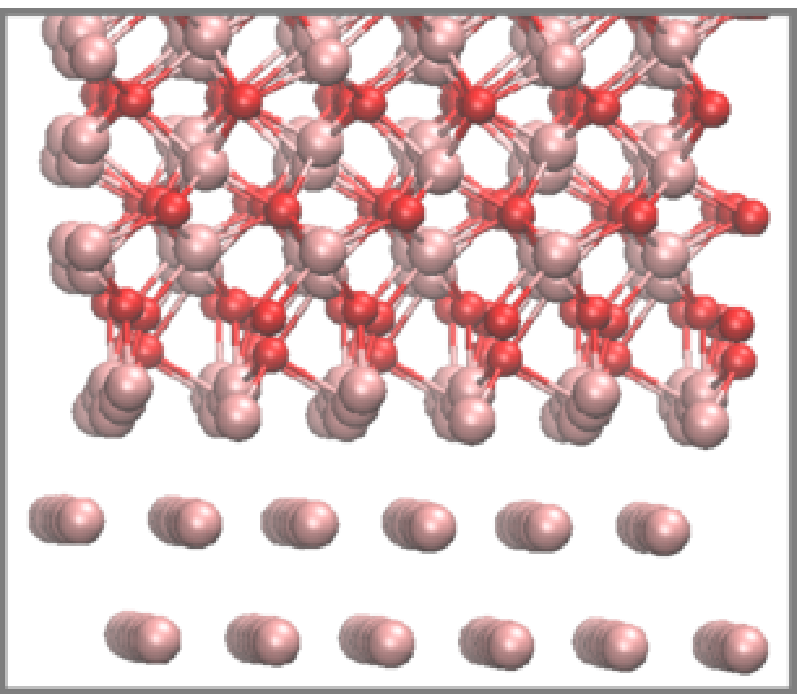} (b) }}
\centerline{\hbox{ (c) \epsfxsize=2.00in \epsfbox{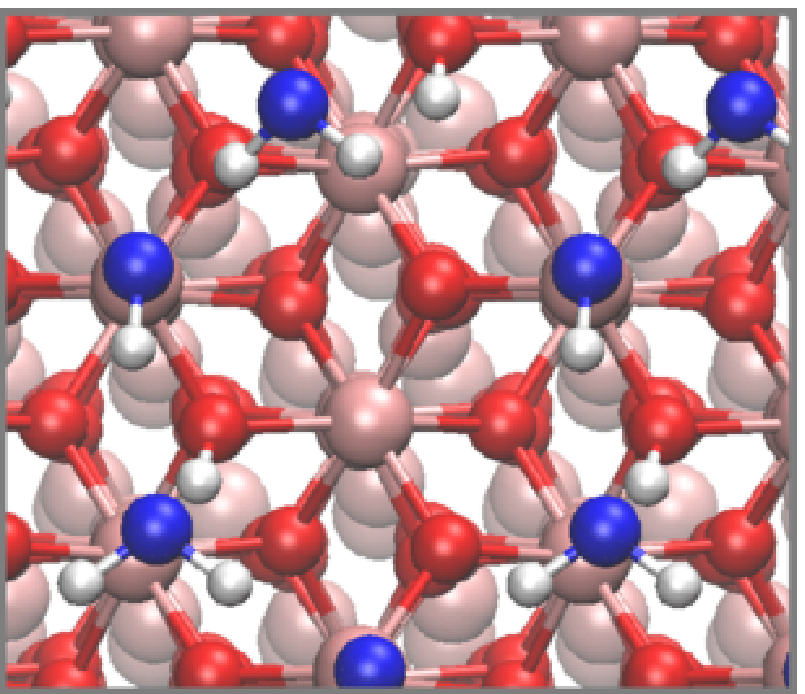} 
		   \epsfxsize=2.00in \epsfbox{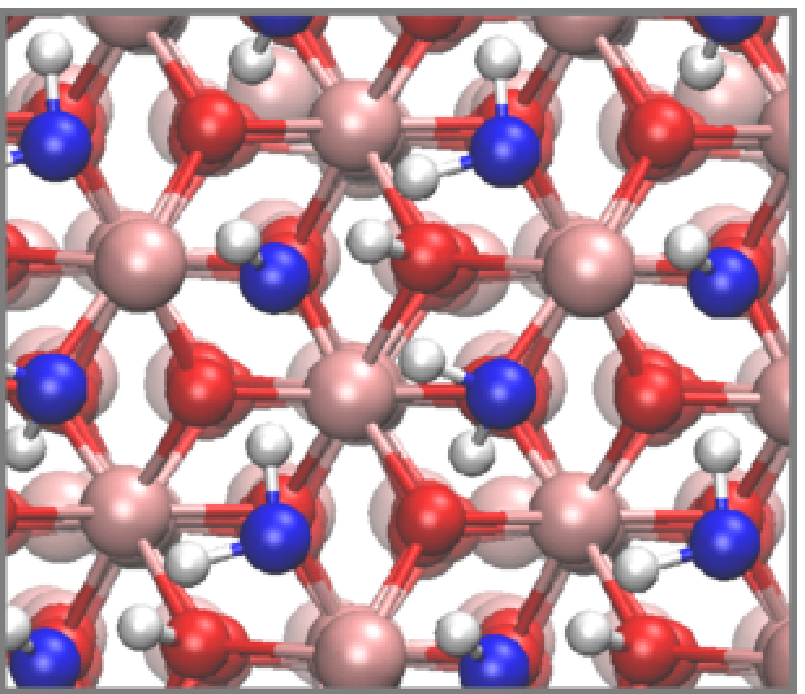} (d) }}
\centerline{\hbox{ (e) \epsfxsize=3.90in \epsfbox{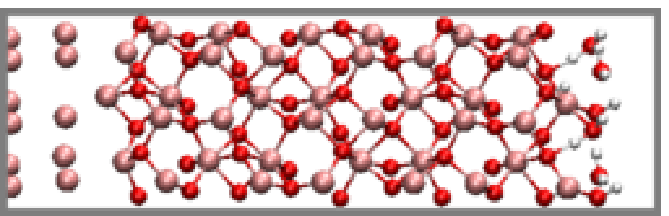} }}
\caption[]
{\label{fig9} \noindent
(a)-(b) Close-up views of two possible Al/$\alpha$-Al$_2$O$_3$ (0001)
with and without AIMD pre-equilibration respectively.
(c)-(d) Close-up views of two possible $\alpha$-Al$_2$O$_3$ (0001) surfaces, 
with 2 or 3 adsorbed H$_2$O per exposed surface Al$^{3+}$ cations.  Some of
these H$_2$O are dissociated.  (e) Simulation cell with a different number
of Al atoms at the Al-metal side of the Al/Al$_2$O$_3$ interface.  The
$\alpha$-Al$_2$O$_3$ (0001)$|$H$_2$O surface in Fig.~\ref{fig9}e in fact
resembles that of Fig.~\ref{fig9}d.
}
\end{figure}

Two interfacial structures at the Al/oxide interface, with and without AIMD
pre-equilibration for 1.5~ps before optimizing the atomic configurations, are
shown in Fig.~\ref{fig9}a-b.  Both start with the ``FCC'' metal/oxide
alignment,\cite{siegel2002} and are created by adding a Al below each
undercoordinated O-anion on the $\alpha$-Al$_2$O$_3$ (0001) inner surface
(Fig.~\ref{fig9}b).  Upon applying AIMD (Fig.~\ref{fig9}a), three of the six
Al atoms per doubled surface unit cell thus added are transferred on to the
Al metal region, and the other three remain bonded to O-anions.  This Al-O
bonding configuration is similar with that in Ref.~\onlinecite{bredas}; it
is slightly different from the optimized configuration using an undoubled
surface unit cell where the equivalent of 4 surface O atoms would remain
bonded to Al below.\cite{siegel2002} The AIMD equilibration lowers the total
energy of the simulation cell, by $\sim$1.7~eV.

Two possible interfacial structures at the oxide surface, with 2- and
3-adsorbed H$_2$O molecules per exposed surface Al$^{3+}$, are depicted in
Fig.~\ref{fig9}c-d.  Our previous work,\cite{pccp}, and the current 
Fig.~\ref{fig2}c, have 2 H$_2$O per surface Al$^{3+}$, while Fig.~\ref{fig2}a
has 3 H$_2$O per surface Al$^{3+}$.  With these two sets of interfaces, we
create three slab configurations: (I) Fig.~\ref{fig9}a \& Fig.~\ref{fig9}c;
(II) Fig.~\ref{fig9}a \& Fig.~\ref{fig9}d (which is effectively 
Fig.~\ref{fig2}a); (III) Fig.~\ref{fig9}b \& Fig.~\ref{fig9}d.  
See Table~\ref{table5}.

\begin{figure}
\centerline{\hbox{ \epsfxsize=4.50in \epsfbox{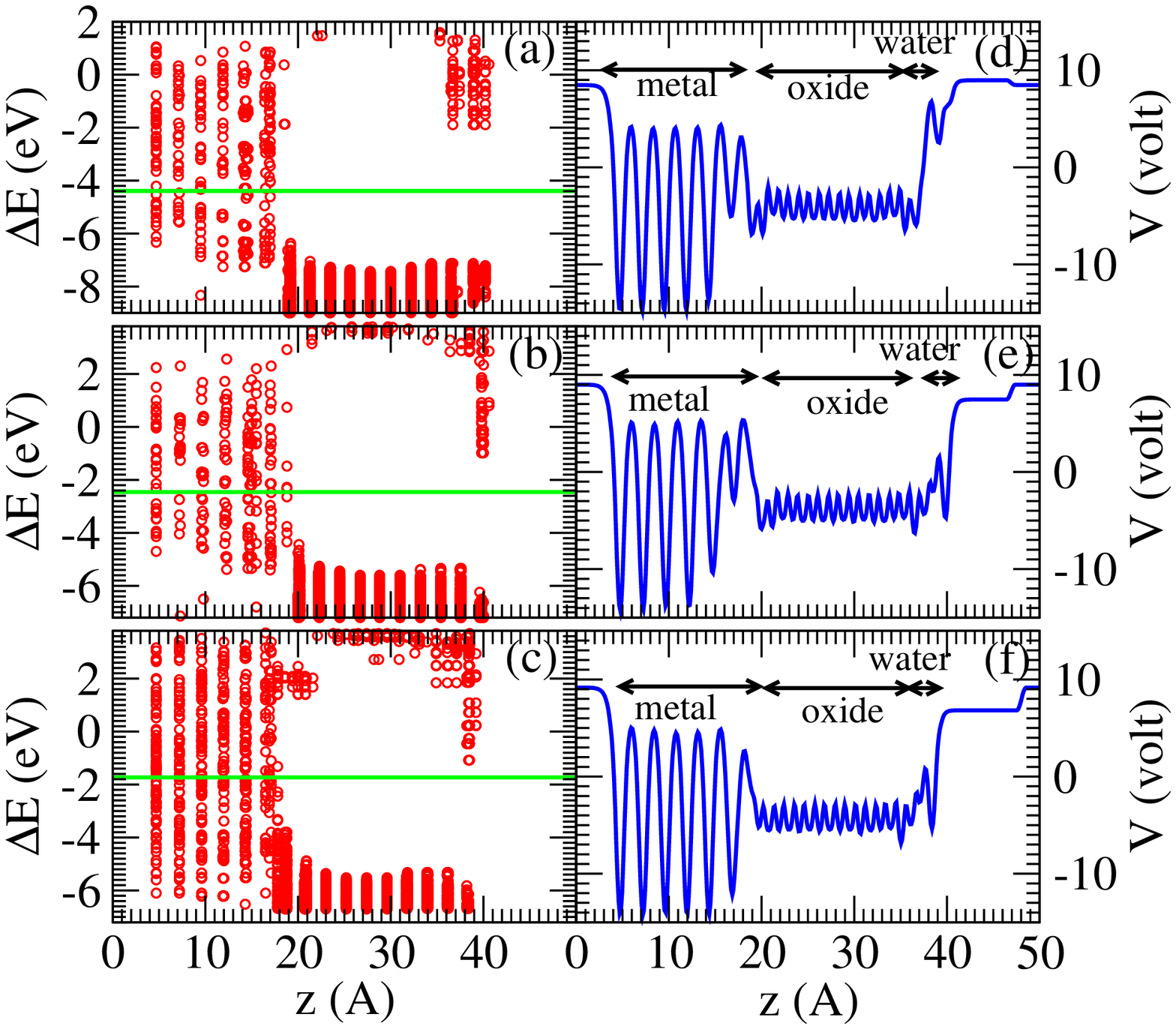} }}
\caption[]
{\label{fig10} \noindent
(a)-(c) Localized densities of state for systems (I)-(III), respectively.
(d)-(f) Electrstatic potentials for systems (I)-(III), respectively.
The green lines denote Fermi levels, and vacuum is at $\Delta E$=0.0~eV.
}
\end{figure}

The corresponding local densities of state and electrostatic potential
along the $z$-direction of these three slabs, after optimization of atomic
coordinates, are depicted in Fig.~\ref{fig10}a-c and Fig.~\ref{fig10}d-f,
respectively.  The overall ${\cal V}_e$ are -0.05~V, -1.98~V, and -2.69~V
vs.~SHE for systems (I)-(III), respectively.  This suggests that both
interfaces affect the overall voltage but the oxide-solvent interface can be
dominant.  Comparing (II) and (III), the AIMD-equilibrated structure contains
fewer Al$^{\delta +}$-O bonds at the metal/oxide, which is correlated with
a smaller overall dipole moment and a more positive voltage via Eq.~\ref{eqdip}.
The variation in ${\cal V}_e$ is larger than in Ref.~\onlinecite{bredas} 
because the Fig.~\ref{fig9}b is metastable by a significant amount of energy.

Fig.~\ref{fig10} depicts local densities of states (LDOS) and integrated
electrostatic potentials along the $z$-direction ($V(z)$). They highlight
how different interfaces affect local LDOS features.  Systems (I) and (II)
exhibit similar LDOS and $V(z)$, except at the oxide-``solvent'' interface
at $z$$\sim$40~\AA, \color{black} (the water zone) \color{black} and except that
the LDOS are globally shifted from each other by $\sim$2~eV.  For example,
The offsets between the \color{black} oxide zone \color{black} VBE and the
Fermi level (i.e., $\Delta E_{\rm FVBE}$) are $\sim$2.88~eV in both
in both (I) and (II).  These systems have the same metal-film
interface, and the global shift must be due to the film-solvent interface
(Table~\ref{table5}).  In contrast, systems (II) and (III) differ
structurally only at the metal-film interface.  There are substantial
differences in the LDOS and $V(z)$ between these systems.  For example,
 $\Delta E_{\rm FVBE}$ now rises to 3.60~eV in (III) compared to 2.88~eV
in (I) and (II).  However, as expected, the LDOS near the film-solvent
interfaces \color{black} (the water zone) \color{black} are very similar
(but slightly shifted in the $z$ direction because of structural changes at
the metal/oxide interface).  

\begin{table}\centering
\begin{tabular}{l|l|l|l|l} \hline
system & m$|$f & f$|$s & ${\cal V}_e$  & $\Delta E_{\rm FVBE}$\\ \hline
(I) & Fig.~\ref{fig9}a & Fig.~\ref{fig9}c & -0.02~V & 2.64 eV \\
(II) & Fig.~\ref{fig9}a & Fig.~\ref{fig9}d & -1.98~V & 2.88 eV \\
(III) & Fig.~\ref{fig9}b & Fig.~\ref{fig9}d & -2.69~V & 3.59 eV \\
alt &  Fig.~\ref{fig9}e & Fig.~\ref{fig9}e & -1.72~V & 2.64 eV \\
\hline
\end{tabular}
\caption[]
{\label{table5} \noindent
Attributes of systems (I)-(III) examined in the appendix.  Voltages are
referenced to SHE.  For the f$|$s interface, Fig.~\ref{fig9}e and 
Fig.~\ref{fig9}d are identical.
}
\end{table}

The system (II), with the AIMD-equilibrated Al (001)$|$$\alpha$-Al$_2$O$_3$
(0001) interface depicted in Fig.~\ref{fig9}d, appears similar to that in
Ref.~\onlinecite{bredas}.  To further check the results of variations in the
number of Al atoms in the top layer of the metal region per doubled surface
cell, we have created yet another Al$|$Al$_2$O$_3$ interfacial simulation cell
with 5 rather than 3 Al atoms on the top metal layer per doubled cell and
the outer oxide surface structure of Fig.~\ref{fig9}d.  The entire system is
depicted in Fig.~\ref{fig9}e.  This alternative configuration (``alt''  in
Table~\ref{table5}) yields a ${\cal V}_e$ within 0.26~V of that of system (II).
The offsets between the \color{black} oxide zone \color{black} VBE and the
Fermi level ($\Delta E_{\rm FVBE}$) are 2.64~eV and 2.88~eV in the two cases,
within 0.24~eV of each other.

From the above analysis, we make the following observations about the
metal/oxide and the oxide/vacuum interfaces.  (i) The metal/oxide interface
qualitatively modifies the orbital alignment across the oxide film.  (ii)
Varying the number of partially charged Al$^{\delta }$ coordinated to the
bottom layer of oxide at the metal/oxide interface has the most significant
effect on the orbital alignment inside the oxide.  (iii) Varying the number
of Al atoms at the top layer of the metal surface at the interface, not
coordinated to O-atoms in the oxide, has smaller
effects on the band structure because such Al atoms are metallic and mostly
uncharged, and do not give rise to large dipole moment changes that
modify the band alignment.  (iv) The structure of the oxide film outer surface
has a significant effect on the overall work function and ${\cal V}_e$.
However, its effect is to exert a global shift in the band structure and does
not strongly affect the orbital alignment inside the oxide film
-- unless this interface is charged (e.g., via creating proton vacancies).

Finally, as discussed in Sec.~\ref{dft}, the interface between Al(111) and
$\alpha$-Al$_2$O$_3$ with a grain boundary is too large for us to perform
AIMD pre-equilibration simulations.  Instead, we displace the metal
and oxide slabs in the $x$- and $y$- directions for a total of 13 cases.
We find that ${\cal V}_e$ averages to -1.47$\pm$0.03~V, in good agreement
with the -1.54~V for the configuration considered in the main text.  The 
simulation cell total energies exhibit a standard deviation 0.04~eV/nm$^2$
in the 1.29~nm$^2$ lateral surface area simulation cell; the LDOS are
qualitatively similar.  Hence we conclude that the lateral registry of the
metal and oxide slabs does not strongly affect the LDOS in this system.

\section*{Acknowledgement}
 
We thank Nancy Missert, Peter Schultz, Quinn Campbell, and Katherine
Jungjohann for useful suggestions.  This work is funded by the Advanced
Strategic Computing (ASC) Program.
Sandia National Laboratories is a multi-mission laboratory managed and operated
by National Technology and Engineering Solutions of Sandia, LLC, a wholly owned
subsidiary of Honeywell International, Inc., for the U.S. Department of
Energy’s National Nuclear Security Administration under contract DE-NA0003525.
This paper describes objective technical results and analysis.  Any subjective
views or opinions that might be expressed in the document do not necessarily
represent the views of the U.S. Department of Energy or the United States
Government.



\end{document}